\documentclass[preprint,onecolumn,superscriptaddress,nofootinbib]{revtex4}

\usepackage{graphicx}
\usepackage{dcolumn}
\usepackage{bm}
\usepackage{amssymb}
\usepackage{amsfonts}
\usepackage{latexsym}
\usepackage{enumerate}
\usepackage[dvipdfmx]{color} 
\usepackage{amsmath}
\usepackage[dvipdfmx]{epsfig}
\usepackage{times}
\usepackage{algorithm}
\usepackage{algorithmic}
\usepackage{cases}

\newcommand{\red}{\color{black}}
\newtheorem{theorem}{Theorem}

\newtheorem{corollary}{Corollary}
\newtheorem{definition}{Definition}

\begin{document}
\widetext
\begin{flushright}
YITP-19-89
\end{flushright}
\title{Sumcheck-based delegation of quantum computing to rational server\footnote{A shortened version of this paper appeared in {\it Proceedings of the 16th International Conference on Theory and Applications of Models of Computation (TAMC 2020)}, vol. 12337 of {\it Lecture Notes in Computer Science}, pp. 69-81, Springer, 2020~\cite{TMT20}.}}
\author{Yuki Takeuchi}
\email{yuki.takeuchi.yt@hco.ntt.co.jp}
\affiliation{NTT Communication Science Laboratories, NTT Corporation, 3-1 Morinosato Wakamiya, Atsugi, Kanagawa 243-0198, Japan}
\author{Tomoyuki Morimae}
\affiliation{Yukawa Institute for Theoretical Physics, Kyoto University, Kitashirakawa Oiwakecho, Sakyoku, Kyoto 606-8502, Japan}
\affiliation{JST, PRESTO, 4-1-8 Honcho, Kawaguchi, Saitama 332-0012, Japan}
\author{Seiichiro Tani}
\affiliation{NTT Communication Science Laboratories, NTT Corporation, 3-1 Morinosato Wakamiya, Atsugi, Kanagawa 243-0198, Japan}

\begin{abstract}
Delegated quantum computing enables a client with weak computational power to delegate quantum computing to a remote quantum server in such a way that the integrity of the server can be efficiently verified by the client.
Recently, a new model of delegated quantum computing has been proposed, namely, rational delegated quantum computing.
In this model, after the client interacts with the server, the client pays a reward to the server depending on the server's messages and the client's random bits.
The rational server sends messages that maximize the expected value of the reward.
It is known that the classical client can delegate universal quantum computing to the rational quantum server in one round.
In this paper, we propose novel one-round rational delegated quantum computing protocols by generalizing the classical rational sumcheck protocol.
An advantage of our protocols is that they are gate-set independent: the construction of the previous rational protocols depends on gate sets, while our sumcheck technique can be easily realized with any local gate set (each of whose elementary gates can be specified with a polynomial number of bits).
Furthermore, as with the previous protocols, our reward function satisfies natural requirements (the reward is non-negative, upper-bounded by a constant, and its maximum expected value is lower-bounded by a constant).
We also discuss the reward gap.
Simply speaking, the reward gap is a minimum loss on the expected value of the server's reward incurred by the server's behavior that makes the client accept an incorrect answer.
The reward gap should therefore be large enough to incentivize the server to behave optimally.
Although our sumcheck-based protocols have only exponentially small reward gaps as in the previous protocols, we show that a constant reward gap can be achieved if two noncommunicating but entangled rational servers are allowed. We also discuss whether a single rational server is sufficient under the (widely believed) assumption that the learning-with-errors problem is hard for polynomial-time quantum computing.
Apart from these results, we show, under a certain condition, the equivalence between {\it rational} and {\it ordinary} delegated quantum computing protocols.
This equivalence then serves as a basis for a reward-gap amplification method.\\

\end{abstract}
\maketitle

\section{Introduction}
\label{I}
\subsection{Background}
It is widely accepted that quantum computing outperforms classical computing in several tasks such as integer factorization~\cite{S97}, approximations of Jones Polynomials~\cite{FKW02,FLW02,AAEL07}, and simulations of quantum systems~\cite{L96,GAN14}.
Due to the superiority of quantum computing, huge experimental efforts have sought to realize larger quantum devices, and devices capable of controlling 12--53 qubits have already been implemented~\cite{LLLKPFXLLBBZL17,FMMHHJPHRBL18,WLHCSLCLFJZLLLP18,WLYZ18,google}. 
As the size of quantum devices increases so does the importance of efficiently verifying whether a constructed quantum device correctly works.
The verification of quantum computing also plays an important role in delegated quantum computing, which enables a client with weak computational power to delegate quantum computing to a remote (potentially malicious) server in such a way that the client can efficiently verify whether the server faithfully computes the delegated problem (i.e., can verify the integrity of the server).

One of the most important open problems in the field of quantum computing is whether a classical client can efficiently delegate universal quantum computing to a quantum server while efficiently verifying the integrity of the server.
In delegated quantum computing, the server may be malicious and may make the client accept an incorrect answer.
Therefore, the client has to efficiently verify the integrity of the server using only classical computation and communication.
Furthermore, the honest server's computational power should be bounded by polynomial-time quantum computing, because delegated quantum computing with a server having unbounded computational power is unrealistic.
This limitation is the large difference between delegated quantum computing and interactive proof systems for {\sf BQP}.
In interactive proof systems, the computational power of the prover (i.e., the server) is unbounded.
Indeed, the known construction of an interactive proof system for {\sf BQP} requires the honest prover to have {\sf PP} computational power~\cite{AG17}, and it is not known whether the honest prover's power can be reduced to {\sf BQP}. Therefore, this open problem cannot be straightforwardly solved from the well-known containment {\sf BQP}$\subseteq${\sf PSPACE}$=${\sf IP}~\cite{S90}.

So far, several partial solutions to this open problem have been obtained.
For example, if small quantum memories, single-qubit state preparations, or single-qubit measurements are allowed for the classical client, the client can efficiently delegate verifiable universal quantum computing to the quantum server~\cite{FK17,ABEM17,HM15,MTH17,FHM18,TM18,TMMMF19}.
As another example, the completely classical client can efficiently delegate it to multiple quantum servers who share entangled states but cannot communicate with each other~\cite{FHM18,RUV13,M16,GKW15,HPF15,CGJV17,NV17,NV18,G19}.
It is also known that some problems in {\sf BQP} can be efficiently verified by interactions between the classical client and quantum server~\cite{M12,DOF18,MTN18,FMNT18}.
Examples of such verifiable problems are integer factorization, recursive Fourier sampling~\cite{M12}, promise problems related to output probability distributions of quantum circuits in the second level of the Fourier hierarchy~\cite{DOF18,MTN18}, and the calculation of the order of solvable groups~\cite{FMNT18}.
Furthermore, it has recently been shown that if the learning-with-errors (LWE) problem is hard for polynomial-time quantum computing, the classical client can delegate verifiable universal quantum computing to a single quantum server whose computational power is bounded by {\sf BQP}\footnote{In this paper, for simplicity, we sometimes use complexity classes (e.g., {\sf BQP} and {\sf BPP}) to represent computational powers. For example, we say that a server (a client) is a {\sf BQP} server (a {\sf BPP} client) when he/she is a polynomial-time quantum server (classical client), i.e., he/she can perform polynomial-time quantum (probabilistic classical) computing.} even in the malicious case~\cite{M18,GV19,CCKW19}.

\begin{figure*}[t]
\includegraphics[width=8cm, clip]{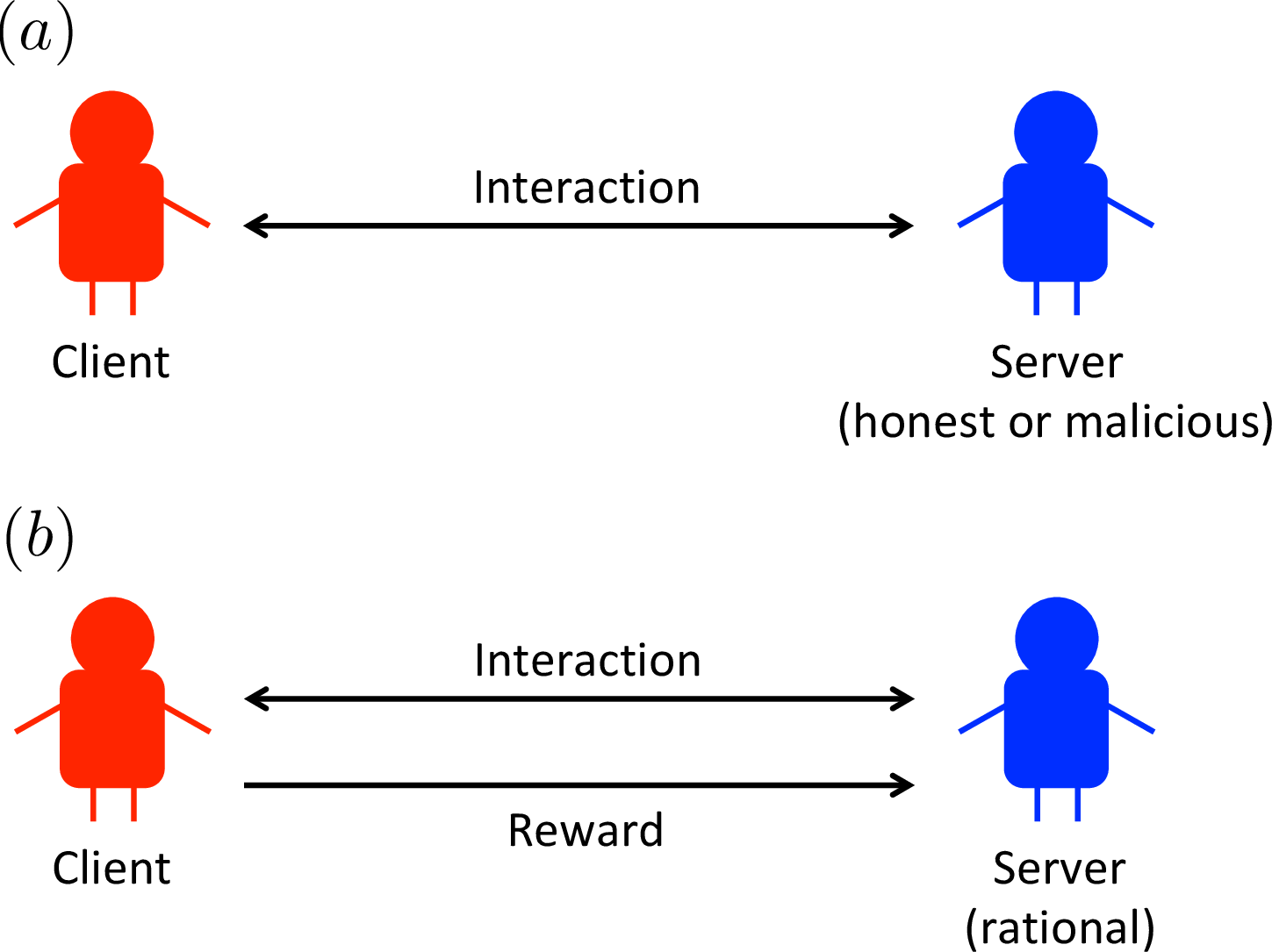}
\caption{Schematics of two types of delegated quantum computing protocols. (a) In ordinary delegated quantum computing, the server may be malicious and try to deceive the client. (b) In rational delegated quantum computing, the client pays a reward to the server after the interaction. The server is always rational, i.e., he/she wants to maximize the expected value of the reward.}
\label{IP}
\end{figure*}

In this paper, we take a different approach to construct protocols for classical-client delegated quantum computing. 
We consider delegating quantum computing to a rational server.
This model was first proposed by Morimae and Nishimura~\cite{MN18} based on the concept of (classical) rational interactive proof systems~\cite{AM12}. 
We note again that the computational power of the server is bounded by {\sf BQP} in rational delegated quantum computing, while it is unbounded in the rational interactive proof systems.
In rational delegated quantum computing, after the client interacts with the server, the client pays a reward to the server depending on the server's messages and the client's random bits. 
As stated above, in {\it ordinary} delegated quantum computing~\cite{FK17,ABEM17,HM15,MTH17,FHM18,TM18,TMMMF19,RUV13,M16,GKW15,HPF15,CGJV17,NV17,NV18,G19,M12,DOF18,MTN18,FMNT18,M18,GV19,CCKW19}, the server may be malicious.
On the other hand, in {\it rational} delegated quantum computing, the server is always rational, i.e., he/she tries to maximize the expected value of the reward (see Fig.~\ref{IP}).
In the real world, there are several situations where service providers want to maximize their profits.
Since rational delegated quantum computing reflects such situations, this model can be considered as another possible situation for delegated quantum computing.
In Ref.~\cite{MN18}, it was shown that the classical client can delegate universal quantum computing to the rational quantum server in one round in such a way that the expected value of the reward attains the maximum when the client obtains the correct answer.

\subsection{Our contribution}
We propose novel one-round delegated quantum computing protocols with a classical client and a rational quantum server.
More precisely, we construct protocols where the classical client can efficiently delegate to the rational quantum server the estimation of output probabilities of $n$-qubit quantum circuits.
Their estimation has many applications such as estimating the expected values of observables, which are quantities interested especially by physicists, and solving decision problems in {\sf BQP}.
Specifically, we consider two classes of quantum circuits: any $n$-qubit polynomial-size quantum circuit with $k$-qubit output measurements, where $k=O(\log{n})$; and approximately $t$-sparse $n$-qubit polynomial-size quantum circuits with $n$-qubit output measurements, where $t$ is a polynomial in $n$.
Here, $t$-sparse means that at most $t$ output probabilities are non-zero (for the formal definition, see Sec.~\ref{QC}).
Since the goal of our rational protocols is to delegate the estimation of the output probabilities, we, for clarity, refer to our protocols as delegated quantum estimating protocols.
Using one of our delegated quantum estimating protocols, we can also construct a one-round rational delegated quantum computing protocol for any decision problem in {\sf BQP}.
Intuitively, using a certain {\sf BQP}-complete problem~\cite{ABEM17}, any {\sf BQP} problem can be reduced to the estimation of the probability of the first qubit being projected onto $|1\rangle$.
Therefore, our argument works.
Note that it is still open whether our results can be generalized to the estimation of output probabilities of any $n$-qubit polynomial-size quantum circuit with $n$-qubit output measurements. 

Our protocols can be applied to a broader class of universal gate sets than the previous protocols~\cite{MN18}.
They work for any universal gate set each of whose elementary gates acts on at most $O(\log{n})$ qubits, while the previous protocols are tailored for Clifford gates plus $T\equiv|0\rangle\langle 0|+e^{i\pi/4}|1\rangle\langle 1|$ or classical gates plus the Hadamard gate.
Note that we only consider gate sets whose elementary gates can be specified with a polynomial number of bits.
As another difference from Ref.~\cite{MN18}, we show that our protocol can be applied to approximately sparse quantum circuits.

{\red Three} conditions should be satisfied {\red by} rational delegated quantum computing protocols:
\begin{itemize}
\item[1.] The reward is upper-bounded by a constant.
\item[2.] The reward is always non-negative if the {\sf BQP} server takes an optimal strategy that maximizes its expected value\footnote{More precisely, the server takes an optimal strategy that can be executed in quantum polynomial time, because we assume that the computational power of the server is bounded by {\sf BQP}. Throughout this paper, the server's optimization is limited to one that can be performed in quantum polynomial time unless explicitly noted otherwise.}.
\item[3.] The maximum of the expected value of the reward is lower-bounded by a constant, where the maximization is taken over {\red all the server's} strategies that can be realized in quantum polynomial time{\red.}
\end{itemize}
The first condition is natural because the client's budget is limited.
The second condition is also natural because a negative reward means that the server pays the reward to the client.
Indeed, the original paper on rational interactive proof systems~\cite{AM12} pointed out that the reward must be non-negative and upper-bounded by a constant.
Furthermore, in Ref.~\cite{AM13}, the non-negativity of the reward (namely, ex-post individual rationality) is listed as one of crucial properties of reward functions.
The third condition guarantees that the server can obtain at least a constant reward on average if the server is rational.

{\red Additionally, the following optional condition would improve the practicality of rational delegated quantum computing protocols:}
\begin{itemize}
{\red \item[4.] The reward gap~\cite{GHRV14} is larger than a constant (or $1/n^{O(1)}$). Here, simply speaking, the reward gap is a minimum loss on the expected value of the server's reward incurred by the server's behavior that makes the client accept an incorrect answer. Note that such behavior may require computational power beyond {\sf BQP}, while we limit the optimal strategy maximizing the expected value to one that can be executed in quantum polynomial time.}
\end{itemize}
Someone may think that the fourth condition is not necessary for practical delegated quantum computing protocols.
In practice, however, there is a possibility where the server selects whether he/she behaves rationally or irrationally depending on the reward gap{\red\footnote{\red From a purely game-theoretic perspective, the server behaves rationally regardless of the value of the reward gap. \red However, it should be hard to model all situations in the real world by using the game theory due to the difficulty of defining an appropriate cost function. For example, in our situation, there may exist a server who wants to send an incorrect answer even if he/she losses a little money but does not want to lose much money.}}.
By satisfying the fourth condition, we may be able to incentivize the server to perform rationally, i.e., to take an optimal strategy that maximizes the expected value of the reward.

The protocols of Ref.~\cite{MN18} and our protocols satisfy only conditions 1--3.
Whether the above four conditions can be satisfied simultaneously is an open problem.
In Ref.~\cite{MN18}, it is shown that if the reward gap is larger than $1/f(n)$ with a polynomial $f(n)$, a super-polynomial increase of the reward (i.e., the violation of the first condition) is unavoidable in one-round protocols with a single server unless {\sf BQP}$\subseteq{\sf \Sigma_3^P}$.
Since this inclusion is considered unlikely given the oracle separation between {\sf BQP} and {\sf PH}~\cite{RT19}, this implies that it may be impossible to satisfy the above four conditions simultaneously in one-round protocols with a single server.
Indeed, all existing rational delegated quantum computing protocols, including ours, are one-round and have only exponentially small gaps under the first three conditions.

In this paper, we also show that a constant reward gap can be achieved if two noncommunicating but entangled servers are allowed.
More precisely, for {\sf BQP} problems, we construct a multi-rational-server delegated quantum computing protocol that satisfies all four conditions simultaneously.
We also discuss whether a single server is sufficient
under the (widely believed) assumption that the LWE problem is hard for polynomial-time quantum computation.
Note that it is still open whether a constant reward gap can be achieved unconditionally.

Apart from these results, we also give, under the certain condition introduced in Ref.~\cite{CMS18}, a relation between {\it rational} and {\it ordinary} delegated quantum computing protocols.
More precisely, we show that under that condition, these two delegated quantum computing protocols can be converted from one to the other and vice versa.
This equivalence may provide a new approach to tackle the open problem of whether a classical client can efficiently delegate universal quantum computing to a (non-rational) quantum server while efficiently verifying the server's integrity.
Based on this equivalence, we give an amplification method for the reward gap.
Under the certain condition, we can amplify the reward gap from $1/2^{n^{O(1)}}$ to a constant.

\subsection{Overview of techniques}
To construct our delegated quantum estimating protocols, we utilize the rational sumcheck protocol~\cite{GHRV16}. The rational sumcheck protocol has been proposed to show that the calculation of $\sum_{i=1}^lx_i$ with integers $x_i\in\{0,\ldots,M-1\}$ can be verified by an $O(polylog(lM))$-time classical verifier in a one-round rational interactive proof system.
To apply the rational sumcheck protocol to our protocols, we generalize it so that it works for complex numbers. 
Then, by combining the generalized rational sumcheck protocol with the Feynman path integral, we construct one-round rational delegated quantum estimating protocols for the two types of quantum circuits, including ones that can solve any {\sf BQP} problem.

Our sumcheck-based protocols work for a broader class of universal gate sets than that used in the previous protocols~\cite{MN18}.
This difference is due to the decomposition method of output probabilities of the delegated quantum circuit. In the previous protocols, the output probability is decomposed using tree structures that are tailored for two specific gate sets--Clifford gates plus the $T$ gate or classical gates plus the Hadamard gate. On the other hand, in our protocol, the output probability is decomposed using the Feynman path integral.

Furthermore, one of our sumcheck-based protocols can be used to delegate the estimation of output probabilities of approximately sparse quantum circuits.
The intuitive reason we can do so is that the maximum of the expected value of the reward monotonically increases as an output probability of the delegated quantum circuit increases.
This implies that the rational server has to send high output probabilities to maximize the expected value of the reward as much as possible.
Since output probabilities of approximately sparse quantum circuits can be approximated by the set of such high output probabilities as shown in Ref.~\cite{SN13}, our rational protocol works.

To construct other rational delegated quantum computing protocols with a constant reward gap, we utilize multiprover interactive proof systems with a constant number of provers and a constant completeness-soundness gap~\cite{FHM18,RUV13,GKW15,CGJV17,NV17,NV18,G19}.
By following a construction used in Ref.~\cite{AM12}, we incorporate the multiprover interactive proof systems into a multi-rational-server delegated quantum computing protocol.
As a result of this construction, the completeness-soundness gap is converted to the reward gap without changing the value of the gap.
More precisely, when the server's computational power is bounded by {\sf BQP}, the obtained reward gap is decreased by a $1/2^{n^{O(1)}}$ additive term from the original completeness-soundness gap. In order to achieve the conversion without changing the value, an unbounded computational power is required.
The same argument can also be applied to Mahadev's single-prover interactive argument system~\cite{M18}, which relies on the hardness assumption of LWE problems. 

Finally, we show that, under the certain condition introduced in Ref.~\cite{CMS18}, {\it rational} and {\it ordinary} delegated quantum computing protocols can be converted from one to the other and vice versa.
To convert ordinary delegated quantum computing protocols to rational ones, we show that the construction in the previous paragraph can be used even under the condition.
For the reverse conversion, we utilize the construction in Ref.~\cite{CMS18}.
Using this conversion between rational and ordinary delegated quantum computing protocols, we show that the amplification of the reward gap can be replaced with that of the completeness-soundness gap under that condition.
This means that the traditional amplification method for the completeness-soundness gap can be used to amplify the reward gap to a constant.
By virtue of the condition, we also show that this reward-gap amplification method works even when the original reward gap is exponentially small\footnote{Note that this does not mean that the traditional amplification method for the completeness-soundness gap works for an exponentially small gap.}.

\section{Preliminaries}
In this section, we give some preliminaries.
In Sec.~\ref{2A}, we define rational delegated quantum computing.
In Sec.~\ref{IIIA}, we define the reward gap.
In Sec.~\ref{QC}, we define approximately sparse quantum circuits.
In Sec.~\ref{2D}, we introduce the {\sf BQP}-complete problem used in this paper.
\subsection{Rational delegated quantum computing}
\label{2A}
In this subsection, we define rational delegated quantum computing.
This definition (Definition~\ref{RDQCRDQC}) will be used in Secs.~\ref{IIIA} and \ref{X}.
To understand Sec.~\ref{II}, it might be possible to skip this subsection.
Following the original definition of rational interactive proof systems~\cite{AM12}, we first define the transcript $\mathcal{T}$, the server's view $\mathcal{S}$, and the client's view $\mathcal{C}$ as follows:
\begin{definition}
\label{RDQC1}
We assume that $k$ is odd.
Given an instance $x$ and a round $i$, we define the $i$th transcript $\mathcal{T}_i$, the $i$th server's view $\mathcal{S}_i$, and the $i$th client's view $\mathcal{C}_i$ as follows $(0\le i\le k)$:
\begin{itemize}
\item $\mathcal{T}_0=\mathcal{S}_0=\mathcal{C}_0=\{x\}$.
\item When $i$ is odd, $\mathcal{T}_i=\{\mathcal{T}_{i-1},a_i\}$, where $a_i$ is the $i$th server's message. On the other hand, when $i(>0)$ is even, $\mathcal{T}_i=\{\mathcal{T}_{i-1},b_i\}$, where $b_i$ is the $i$th client's message.
\item For odd $i$, $\mathcal{S}_i=\{\mathcal{S}_{i-2},\mathcal{T}_{i-1},V_i\}$, where $V_i$ is a quantum circuit used to compute $a_i$. Note that $\mathcal{S}_i$ and $V_i$ are not defined for even $i$ because the even-numbered round is a communication from the client to the server.
\item For even $i$, $\mathcal{C}_i=\{\mathcal{C}_{i-2},\mathcal{T}_{i-1},r_i\}$, where $r_i$ is a random bit string used to compute $b_i$. Note that $\mathcal{C}_i$ is not defined for odd $i$ because the odd-numbered round is a communication from the server to the client.
\end{itemize}
For all $i$, messages $a_i$ and $b_i$ are polynomial lengths.
Particularly, $b_i$ is generated from  $\mathcal{C}_i$ in classical polynomial time.
The quantum circuit $V_i$ is decided from $\mathcal{S}_{i-2}$.
\end{definition}

Based on Definition~\ref{RDQC1}, we define the following interaction consisting of $k$ rounds between a {\sf BPP} client and a server, where we call it $k$-round interaction:
\begin{definition}
\label{RDQC}
Let $k$ be odd.
This means that the protocol begins with the server's step. When $k$ is even, the following definition can be adopted by adding a communication from the server to the client at the beginning of the protocol.
Let us consider the following $k$-round interaction:
\begin{enumerate}
\item A {\sf BPP} client interacts with a server $k$ times. In the $i$th round for odd $i$, the server sends $a_i$ to the client. In the $i$th round for even $i$, the client sends $b_i$ to the server.
\item The client efficiently calculates a predicate on the instance $x$ and the $k$th transcript $\mathcal{T}_k$. If the predicate evaluates to $o=1$, the client answers YES. On the other hand, if $o=0$, the client answers NO.
\item The client efficiently calculates the reward{\red\footnote{\red Here, we define the reward as a real value. It should be approximated by a polynomial-size bit string in actual protocols. In our results, the reward gap (see Definition~\ref{rgap}) can be at least $e^{-f(n)}$ for a polynomial $f(n)$ in the problem size $n$. Therefore, the approximation does not affect our results. Note that if the reward gap is a constant, a constant number of bits are sufficient to approximate the reward.}} $R\in[0,c]$ and pays it to the server, where $c$ is a positive constant. Note that it is not necessary for the client and server to know the value of $c$\footnote{There may exist functions such that their upper bounds can be efficiently shown to be constant but their exact values cannot be efficiently derived. Such functions can also be used as reward functions.}. The reward function $R:\{0,1\}^\ast\times\{0,1\}^{poly(|x|)}\times\{0,1\}^{poly(|x|)}\rightarrow\mathbb{R}_{\ge 0}$ depends on the instance $x\in\{0,1\}^\ast$, the $k$th transcript $\mathcal{T}_k\in\{0,1\}^{poly(|x|)}$, and the client's random bits $r_{k+1}\in\{0,1\}^{poly(|x|)}$.
\end{enumerate}
\end{definition}

Rational delegated quantum computing for decision problems is defined as follows:
\begin{definition}
\label{RDQCRDQC}
{\red Let $\mathbb{E}[f]$ denote the expectation value of a function $f$.
Let $\mathcal{D}_k$ be a distribution that the $k$th transcript follows.}
The $k$-round interaction defined in Definition~\ref{RDQC} is called a $k$-round rational delegated quantum computing protocol for decision problems if the following conditions {\red (Eqs.~(\ref{RDQCyes}) -- (\ref{cyesno}))} hold: {\red for} a language $L\subseteq\{0,1\}^\ast$ in {\sf BQP}, if $x\in L$, there exists a distribution $\mathcal{D}_{\rm YES}$ that can be generated in quantum polynomial time, such that
\begin{eqnarray}
\label{RDQCyes}
{\rm Pr}[o=1\ |\ \mathcal{D}_k=\mathcal{D}_{\rm YES}]\ge\cfrac{2}{3}
\end{eqnarray}
and
\begin{eqnarray}
\label{cyes}
\mathbb{E}_{\mathcal{T}_k\sim\mathcal{D}_{\rm YES},r_{k+1}}[R(x,\mathcal{T}_k,r_{k+1})]\ge c_{\rm YES}
\end{eqnarray}
with some positive constant $c_{\rm YES}\le c$.
Here, the expectation is taken over the client's random bits $r_{k+1}$ and the distribution $\mathcal{D}_{\rm YES}$ of the $k$th transcript $\mathcal{T}_k$.

On the other hand, if $x\notin L$, there exists a distribution $\mathcal{D}_{\rm NO}$ that can be generated in quantum polynomial time, such that
\begin{eqnarray}
\label{RDQCno}
{\rm Pr}[o=0\ |\ \mathcal{D}_k=\mathcal{D}_{\rm NO}]\ge\cfrac{2}{3}
\end{eqnarray}
and
\begin{eqnarray}
\label{cno}
\mathbb{E}_{\mathcal{T}_k\sim\mathcal{D}_{\rm NO},r_{k+1}}[R(x,\mathcal{T}_k,r_{k+1})]\ge c_{\rm NO}
\end{eqnarray}
with some positive constant $c_{\rm NO}\le c$.
Here, the expectation is taken over the client's random bits $r_{k+1}$ and the distribution $\mathcal{D}_{\rm NO}$ of the $k$th transcript $\mathcal{T}_k$.

To generate distributions $\mathcal{D}_{\rm YES}$ and $\mathcal{D}_{\rm NO}$, the server decides the $i$th message $a_i$ following a distribution
{\red\begin{eqnarray*}
\mathcal{D}_i\equiv\{|\langle a_i|V_i|\psi_{i-2}\rangle|^2\}_{a_i},
\end{eqnarray*}}
where $|\psi_{i-2}\rangle$ is the server's quantum state immediately after generating the $(i-2)$th message $a_{i-2}$ for $i\ge 3$ and is the tensor product of a polynomial number of $|0\rangle$'s for $i=1$, and $V_i$ is a polynomial-time generated quantum circuit such that
\begin{eqnarray}
\label{cyesno}
V_i={\rm argmax}_V\mathbb{E}_{\mathcal{D}_k,\mathcal{T}_k~\sim\mathcal{D}_k,r_{k+1}}[R(x,\mathcal{T}_k,r_{k+1})|\mathcal{S}_{i-2},\mathcal{T}_{i-1},V|\psi_{i-2}\rangle].
\end{eqnarray}
The expectation is taken over $\mathcal{T}_k$, $r_{k+1}$, and all possible distributions $\mathcal{D}_k$ that can be generated in quantum polynomial time and are compatible with the server's view $\mathcal{S}_{i-2}$, the transcript $\mathcal{T}_{i-1}$, and a quantum state $V|\psi_{i-2}\rangle$.
Here, we consider only the maximizations that can be performed in quantum polynomial time.
\end{definition}
Since the server's computational power is bounded by {\sf BQP}, it is in general hard for the server to select an optimal message that satisfies Eqs.~(\ref{RDQCyes}) and (\ref{cyes}).
Therefore, the server's message $a_i$ should be probabilistically generated.
That is why we consider the distribution $\mathcal{D}_{\rm YES}$.
The same argument holds for the NO case.

The value $2/3$ in Eqs.~(\ref{RDQCyes}) and (\ref{RDQCno}) can be amplified to $1-2^{-f(|x|)}$, where $f(|x|)$ is any polynomial in $|x|$, using the standard amplification method (i.e., by repeating steps 1 and 2, and then taking the majority vote among outputs in step 2).
We here mention that the above definition of rational delegated quantum computing protocols satisfies conditions 1--3 in Sec.~\ref{I}.
This is straightforward from $R\in[0,c]$ and Eqs.~(\ref{cyes}) and (\ref{cno}).

The server would like to generate the $i$th message $a_i$ following a distribution that maximizes the expected value of the finally obtained reward.
However, at that time, the server cannot predict the future distribution $\mathcal{D}_k$.
Therefore, the server also takes the expectation over all possible distributions $\mathcal{D}_k$.
The distribution $\mathcal{D}_i$ in Eq.~(\ref{cyesno}) is a distribution that maximizes such expected reward.

All of our rational protocols except for those in Secs.~\ref{IIA} and \ref{IIC} are in accordance with Definition~\ref{RDQCRDQC}.
Our rational protocols in Secs.~\ref{IIA} and \ref{IIC} are rational delegated quantum computing protocols for function problems, which can be defined in a similar way.

For convenience, we define a strategy $s$ as a set of the server's messages $\{a_i\}_i$, which may be adaptively decided according to the previous client's messages.
When we focus on the dependence on the server's messages, we write $\mathbb{E}_{\mathcal{T}_k\sim\mathcal{D},r_{k+1}}[R(x,\mathcal{T}_k,r_{k+1})]$ by $\mathbb{E}_{s\sim\mathcal{D}'}[R(x,s)]$ for short.

\subsection{Reward gap}
\label{IIIA}
Guo {\it et al.} have introduced the reward gap~\cite{GHRV14}, which is also called the utility gap~\cite{CMS16,CMS18}.
For decision problems, the reward gap is defined as follows:
\begin{definition}
\label{rgap}
Let a strategy $s$ be defined as a set $\{a_i\}_{i}$ of the server's messages.
Let $\mathcal{D}$ be a distribution that the server's strategy $s$ follows.
Let $\mathcal{D}_{\rm max}$ be the distribution $\mathcal{D}$, where each message $a_i$ follows the distribution in Eq.~(\ref{cyesno}).
We say that a rational delegated quantum computing protocol has a $1/\gamma(|x|)$-reward gap if for any input $x$,
\begin{eqnarray*}
\mathbb{E}_{s\sim\mathcal{D}_{\rm max}}[R(x,s)]-{\rm max}_{s\in S_{\rm incorrect}}\mathbb{E}[R(x,s)]\ge \cfrac{1}{\gamma(|x|)},
\end{eqnarray*}
where $\gamma(|x|)$ is any function of $|x|$, and $S_{\rm incorrect}$ is the set of the server's strategies that make the client output an incorrect answer with unit probability\footnote{Even if we replace ``unit probability" with ``high probability" such as $1-1/2^{n^{O(1)}}$ and $1-1/n^{O(1)}$, the reward gap in Protocol 2 is still constant.}.
Here, the expectation is also taken over the client's random bits, and the server's strategy $s$ may be adaptively decided according to the client's messages.
Note that $S_{\rm incorrect}$ may include strategies that cannot be executed in quantum polynomial time.
\end{definition}
From Definition~\ref{RDQCRDQC}, if the server's strategy $s$ follows the distribution $\mathcal{D}_{\rm max}$, the client outputs a correct answer with high probability.
$\mathbb{E}_{s\sim\mathcal{D}_{\rm max}}[R(x,s)]$ is the maximum expected value of the reward paid to the rational {\sf BQP} server.
On the other hand, ${\rm max}_{s\in S_{\rm incorrect}}\mathbb{E}[R(x,s)]$ is the maximum expected value of the reward paid to the {\it malicious} computationally-unbounded server if the server wants to maximize the expected value as much as possible while deceiving the client.
This is because the client outputs an incorrect answer when the server takes the strategy $s\in S_{\rm incorrect}$.
As a result, the reward gap represents how much benefit the rational server can obtain compared with the malicious one.

For function problems, we can define the reward gap in a similar way.

\subsection{Approximately sparse quantum circuits}
\label{QC}
An $n$-qubit quantum circuit $U$ consists of elementary gates in a universal gate set.
In this paper, the quantum circuit $U$ is denoted as $U=u_Lu_{L-1}\ldots u_1\equiv\prod_{i=L}^1u_i$, where $u_i$ is an elementary gate in the universal gate set for all $i$, and $L$ is a polynomial in $n$. 
For instance, when we consider $\{CNOT,H,T\}$ as a universal gate set, $u_i$ is the controlled-NOT gate $CNOT$, the Hadamard gate $H$, or the T gate $T$. 
Our argument can be applied to any universal gate set each of whose elementary gates acts on at most $O(\log{n})$ qubits.
Note that each elementary gate is assumed to be specified with a polynomial number of bits.

By using the above notation, $\epsilon$-approximately $t$-sparse polynomial-size quantum circuits are defined as follows:
\begin{definition}[$\epsilon$-approximately $t$-sparse polynomial-size quantum circuit~\cite{SN13}]
\label{sparse}
Consider an $n$-qubit quantum circuit $U\equiv\prod_{i=L}^1 u_i$ with input $|0^n\rangle$, where $L$ is a polynomial in $n$, and each $u_i$ is a unitary gate chosen from a certain universal gate set each of whose elementary gates can be specified with a polynomial number of bits.
Let $q_z\equiv|\langle z|U|0^n\rangle|^2$ be the probability of the quantum circuit outputting $z\in\{0,1\}^n$.
The quantum circuit $U$ is called $\epsilon$-approximately $t$-sparse if there exists a $t$-sparse vector ${\bm v}=(v_z : z\in\{0,1\}^n)$ such that $\sum_{z\in\{0,1\}^n}|q_z-v_z|\le\epsilon$, where a vector is called $t$-sparse if at most $t$ of its coordinates are non-zero.
\end{definition}
Note that in this paper, we assume that $t=f(n)$ and $\epsilon=1/g(n)\ {\red(\le1/6)}$ for some polynomials $f(n)$ and $g(n)$.

\subsection{{\sf BQP}-complete problem}
\label{2D}
Among several {\sf BQP}-complete problems~\cite{KL01,FKW02,FLW02,AAEL07,JW06,AJKR10,MFI14,ABEM17,MTN18}, in this paper, we use the following promise problem:
\begin{definition}[Q-CIRCUIT~\cite{ABEM17}]
\label{qcircuit}
The input is a classical description of an $n$-qubit quantum circuit $U=\prod_{i=L}^1u_i$, where $u_i$ is chosen from a certain universal gate set each of whose elementary gates can be specified with a polynomial number of bits, and $L$ is a polynomial in $n$. If
\begin{eqnarray}
\label{YES}
\langle 0^n|U^\dag(|1\rangle\langle 1|\otimes I^{\otimes n-1})U|0^n\rangle\ge\cfrac{2}{3},
\end{eqnarray}
output YES. On the other hand, if
\begin{eqnarray}
\label{NO}
\langle 0^n|U^\dag(|1\rangle\langle 1|\otimes I^{\otimes n-1})U|0^n\rangle\le\cfrac{1}{3},
\end{eqnarray}
output NO.
It is guaranteed that the input unitary $U$ always satisfies either Eq.~(\ref{YES}) or (\ref{NO}).
\end{definition}

\section{Sumcheck-based rational delegated quantum computing}
\label{II}
In this section, we construct two rational delegated quantum computing protocols for estimating output probabilities of $n$-qubit quantum circuits, which we call the rational delegated quantum estimating protocols. 
In Sec.~\ref{IIA}, we consider any $n$-qubit polynomial-size quantum circuit with $O(\log{n})$-qubit output measurements.
We also show that our protocol satisfies conditions 1--3 mentioned in Sec.~\ref{I}. 
In Sec.~\ref{IIB}, we show that our protocol proposed in Sec.~\ref{IIA} can be applied to classically delegate any decision problem in {\sf BQP}.
In Sec.~\ref{IIC}, we construct another protocol for approximately $t$-sparse $n$-qubit polynomial-size quantum circuits with $n$-qubit output measurements, where $t$ is a polynomial in $n$.

\subsection{Estimating output probabilities of quantum circuits with a logarithmic number of output qubits}
\label{IIA}
In this subsection, we consider an $n$-qubit polynomial-size quantum circuit $U$ with $k=O(\log{n})$ output qubits. 
Let $\{q_z\}_{z\in\{0,1\}^k}$ be the output probability distribution of the quantum circuit $U$, where
\begin{eqnarray*}
q_z\equiv \langle 0^n|U^\dag(|z\rangle\langle z|\otimes I^{\otimes n-k})U|0^n\rangle
\end{eqnarray*}
and $I$ is the two-dimensional identity operator.
We show that if the quantum server is rational, the classical client can efficiently obtain the estimated values $\{p_z\}_{z\in\{0,1\}^k}$ with high probability such that $|p_z-q_z|\le1/f(n)$ for any $z$ and any polynomial $f(n)$.
Therefore, for example, the classical client can approximately sample with high probability in polynomial time from the output probability distribution $\{q_z\}_{z\in\{0,1\}^k}$ of the quantum circuit $U$. 
Before proposing our rational delegated quantum estimating protocol, we calculate $q_z$ using the Feynman path integral. 
Let $U=\prod_{i=L}^1u_i$, where $u_i$ is an elementary gate in a universal gate set for all $i${\red ,} $L$ {\red be} a polynomial in $n${\red , $s^{(i)}$ be an $n$ bit string for $1\le i\le 2L-1$ except for $i=L$, and $s^{(L)}$ be an $n-k$ bit string.
For simplicity, we also define $s$ as a shorthand notation of the $(2L-1)n-k$ bit string $s^{(1)}s^{(2)}\ldots s^{(2L-1)}$.}
The probability $q_z$ is calculated as follows:
\begin{eqnarray}
\nonumber
q_z&=&\langle 0^n|U^\dag(|z\rangle\langle z|\otimes I^{\otimes n-k})U|0^n\rangle\\
\nonumber
&=&\langle 0^n|\left(\prod_{j= L}^1u_j\right)^\dag(|z\rangle\langle z|\otimes I^{\otimes n-k})\left(\prod_{i=L}^1u_i\right)|0^n\rangle\\
\nonumber
&=&\langle 0^n|u_1^\dag\left[\prod_{j=L}^2u_j\left(\sum_{s^{(j-1)}\in\{0,1\}^n}|s^{(j-1)}\rangle\langle s^{(j-1)}|\right)\right]^\dag\left[|z\rangle\langle z|\otimes \left(\sum_{s^{(L)}\in\{0,1\}^{n-k}}|s^{(L)}\rangle\langle s^{(L)}|\right)\right]\\
\label{qz}
&&\left[\prod_{i=L}^2u_i\left(\sum_{s^{(L+i-1)}\in\{0,1\}^n}|s^{(L+i-1)}\rangle\langle s^{(L+i-1)}|\right)\right]u_1|0^n\rangle.
\end{eqnarray}
Here, we define
\begin{eqnarray}
\label{gs}
g(z,s)\equiv\langle 0^n|u_1^\dag\left(\prod_{j=L}^2u_j|s^{(j-1)}\rangle\langle s^{(j-1)}|\right)^\dag |zs^{(L)}\rangle\langle zs^{(L)}|\left(\prod_{i=L}^2u_i|s^{(L+i-1)}\rangle\langle s^{(L+i-1)}|\right)u_1|0^n\rangle{\red .}
\end{eqnarray}
From Eqs.~(\ref{qz}) and (\ref{gs}),
\begin{eqnarray}
\label{qzg}
q_z=\sum_{s\in\{0,1\}^{(2L-1)n-k}}g(z,s).
\end{eqnarray}
As an important point, given $z$ and $s$, the function $g(z,s)$ can be calculated in classical polynomial time.
This is because each elementary gate acts on at most $O(\log{n})$ qubits. 
Note that since there are exponentially many terms in Eq.~(\ref{qzg}), this fact does not contradict the {\sf $\#$P}-hardness of calculating output probabilities of quantum circuits~\cite{FGHP99}. 
Furthermore, from Eq.~(\ref{gs}), the absolute value $|g(z,s)|$ is upper-bounded by $1$. 
Therefore, $0\le (1+{\rm Re}[g(z,s)])/2\le 1$, where ${\rm Re}[g(z,s)]$ is the real part of $g(z,s)$.

\begin{figure*}[t]
\includegraphics[width=8cm, clip]{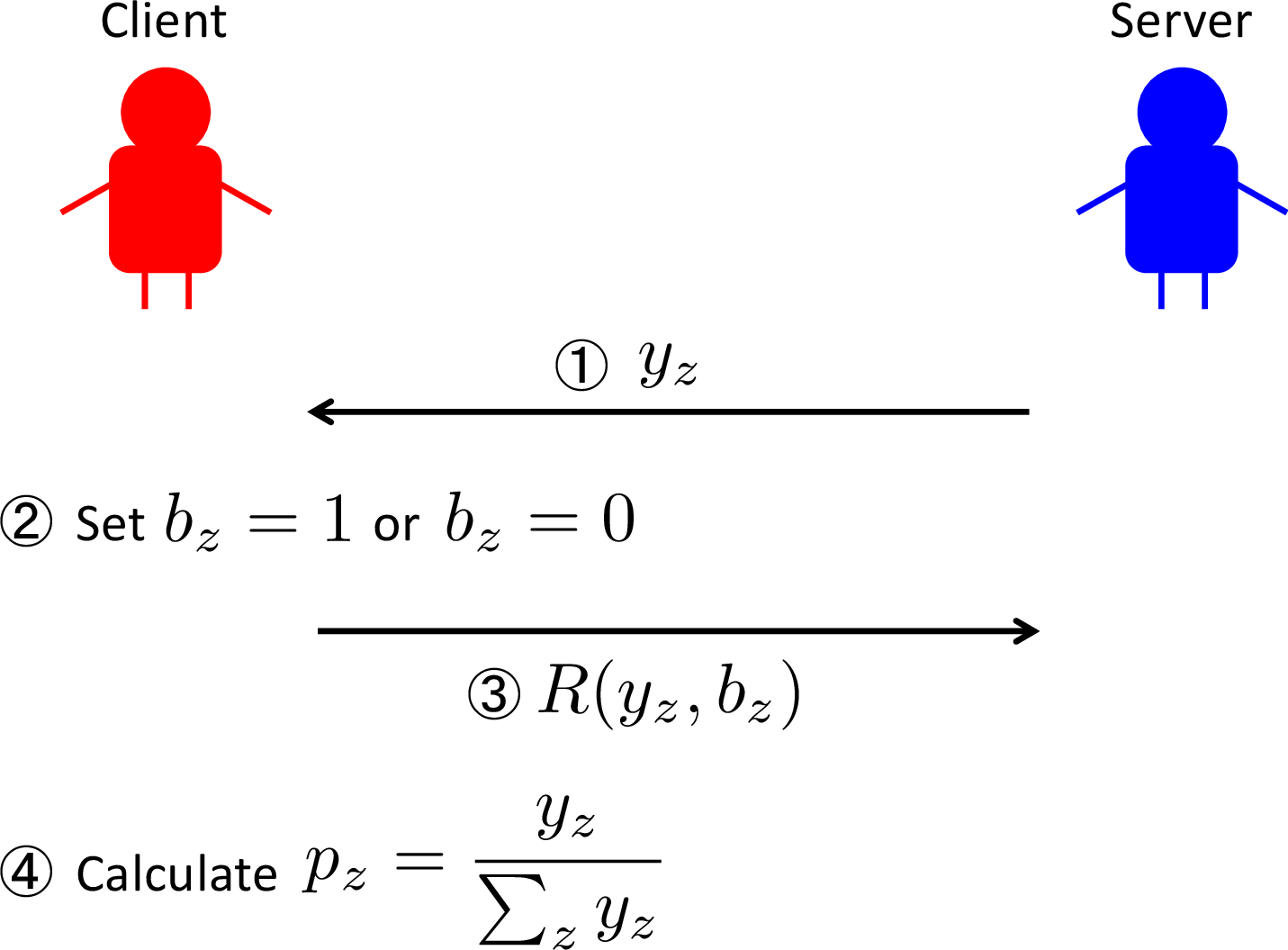}
\caption{The schematic diagram of Protocol 1. First, the server sends $y_z$ to the client in step 1 (a). Second, in steps 1 (b) and (c), the client chooses $b_z=1$ or $b_z=0$ independently of $y_z$. Third, in step 1 (d), the client pays the reward $R(y_z,b_z)$ to the server depending on $y_z$ and $b_z$. The client and the server repeat these procedures for all $z$. Finally, the client calculates $\{p_z\}_z$ from $\{y_z\}_z$ and accepts them as approximations of $\{q_z\}_z$.}
\label{protocolfig}
\end{figure*}

To construct our rational delegated quantum estimating protocol, we use the rational sumcheck protocol~\cite{GHRV16}.
Let us consider the following problem: given $l$ non-negative integers $x_1,\ldots,x_l\in\{0,\ldots,M-1\}$, where $M$ is a positive integer, output $\sum_{i=1}^lx_i$.
The rational sumcheck protocol enables the client to efficiently delegate this problem to the rational server.
The non-negativity of $x_i$ is necessary because the client is required to generate the probability distribution $\{x_i/M,1-x_i/M\}$.
To fit the rational sumcheck protocol to our case, we generalize it for the case of the complex number $x_i$. As a result, we can set $x_i=g(z,s)$ and $z$ to be a certain fixed value. Our protocol runs as follows (see Fig.~\ref{protocolfig}):\\
\noindent
[{\bf Protocol 1}]\\
This protocol depends on the number $k$ of measured qubits, a quantum circuit $U$ to be executed by the server, the number $L$ of elementary gates used in $U$, the number $n$ of input qubits, the required precision characterized by $f(n)$, and the required confidence characterized by $h(n)$ (see Theorem~\ref{sample}).
\begin{enumerate}
\item For all $z\in\{0,1\}^k$, the rational server and the client perform the following steps:
\begin{enumerate}
\item The rational server sends to the client a non-negative real number $y_z$, which is explained later. (Note that $y_z$ is represented by a bit string with logarithmic length; therefore, the message size from the server to the client is logarithmic.)
\item The client samples $s$ uniformly at random from $\{0,1\}^{(2L-1)n-k}$.
\item The client flips a coin that lands heads with probability $(1+{\rm Re}[g(z,s)])/2$. If the coin lands heads, the client sets $b_z=1$; otherwise, $b_z=0$.
\item The client calculates the reward
\begin{eqnarray*}
R(y_z,b_z)\equiv\cfrac{1}{2^k}\Bigg[&&2\cfrac{y_z+2^{(2L-1)n-(k+1)}}{2^{(2L-1)n-k}}b_z+2\left(1-\cfrac{y_z+2^{(2L-1)n-(k+1)}}{2^{(2L-1)n-k}}\right)(1-b_z)\\
&&-\left(\cfrac{y_z+2^{(2L-1)n-(k+1)}}{2^{(2L-1)n-k}}\right)^2-\left(1-\cfrac{y_z+2^{(2L-1)n-(k+1)}}{2^{(2L-1)n-k}}\right)^2+1\Bigg],
\end{eqnarray*}
which is the (slightly modified) Brier's scoring rule~\cite{B50}.
This scoring rule guarantees that the expected value of the reward is maximized when $y_z$ is equal to the probability of $b_z=1$ up to additive and multiplicative factors, which means that $y_z=a {\rm Pr}[b_z=1]+b$ with certain parameters $a$ and $b$.
Then, the client pays the reward $R(y_z,b_z)$ to the rational server.
\end{enumerate}

\item The client calculates
\begin{eqnarray}
\label{step6}
p_z\equiv\cfrac{y_z}{\sum_{z\in\{0,1\}^k}y_z}
\end{eqnarray}
for all $z$.

\end{enumerate}
Since the sampling in step (c) can be approximately performed in classical polynomial time as shown in Appendix A, what the client has to do is simply efficient classical computing.
Furthermore, since the repetitions in step 1 can be performed in parallel, this is a one-round protocol.
Note that except for the communication required to pay the reward to the server, Protocol 1 only requires one-way communication from the server to the client.
Since the functions $f$ and $h$ are parameters of Protocol 1, and $k$, $n$, and a classical description of $U$ are inputs to Protocol 1, we also do not count the communication required to tell them from the client to the server.

We show that $p_z$ satisfies $\sum_{z\in\{0,1\}^k}|p_z-q_z|\le 1/f(n)$ for any fixed polynomial $f(n)$ with high probability.
This means that $p_z$ is an approximated value of $q_z$ for each $z$ with high probability.
More precisely, we show the following theorem:
\begin{theorem}
\label{sample}
Let $f(n)$ and $h(n)$ be any polynomials in $n$.
Let $q_z=\langle 0^n|U^\dag(|z\rangle\langle z|\otimes I^{\otimes n-k})U|0^n\rangle$, and $p_z$ be the probability given in Eq.~(\ref{step6}).
Then, given inputs $k$, $n$, and a classical description of $U$, Protocol 1 with parameters $f(n)$ and $h(n)$ satisfies
\begin{eqnarray}
\label{accuracy}
\sum_{z\in\{0,1\}^k}|p_z-q_z|\le \cfrac{1}{f(n)}
\end{eqnarray}
with probability of at least $1-e^{-h(n)}$.
\end{theorem}
{\it Proof.}
Before the formal proof, we explain our intuitive idea of the proof.
The client's purpose is to obtain $q_z$ (or its approximation) with high probability.
If $y_z=aq_z$ for a constant $a$, this purpose is achieved.
This is because the client obtains $p_z=q_z$ from Eq.~(\ref{step6}).
Therefore, the goal of this proof is to show that $aq_z$ maximizes the expected value of the reward.
(Remember that the rational server sends $y_z$ maximizing the expected value.)
By directly calculating the expected value of the reward $R(y_z,b_z)$ over $b_z\in\{0,1\}$, we can show that the expected value is proportional to $-(y_z-q_z/2)^2$ except for terms independent of $y_z$.
This implies that $y_z=q_z/2$ (the case of $a=1/2$) maximizes the expected value.

The formal proof is as follows.
First, we derive the value of $y_z$ that maximizes the expected value of the reward $R(y_z,b_z)$ over $b_z\in\{0,1\}$.
In this proof, for simplicity, we assume that the sampling of $b_z$ in step (c) can be exactly performed using the method in Appendix A (for the approximation case, see Appendix B).
Let $Y_z\equiv (y_z+2^{(2L-1)n-(k+1)})/2^{(2L-1)n-k}$.
The expected reward when the server sends $y_z$ is
\begin{eqnarray}
\nonumber
&&\mathbb{E}_{b_z}[R(y_z,b_z)]\\
\nonumber
&=&\cfrac{1}{2^k}\sum_{s\in\{0,1\}^{(2L-1)n-k}}\cfrac{1}{2^{(2L-1)n-k}}\Bigg\{\cfrac{1+{\rm Re}[g(z,s)]}{2}\left[2Y_z-Y_z^2-\left(1-Y_z\right)^2+1\right]\\
\nonumber
&&+\cfrac{1-{\rm Re}[g(z,s)]}{2}\left[2\left(1-Y_z\right)-Y_z^2-\left(1-Y_z\right)^2+1\right]\Bigg\}\\
\nonumber
&=&\cfrac{1}{2^k}\left[\cfrac{q_zY_z}{2^{(2L-1)n-k-1}}+1-\cfrac{q_z}{2^{(2L-1)n-k}}-Y_z^2-\left(1-Y_z\right)^2+1\right]\\
\nonumber
&=&\cfrac{1}{2^k}\left[-2\left(Y_z-\cfrac{q_z}{2^{(2L-1)n-k+1}}-\cfrac{1}{2}\right)^2+2\left(\cfrac{q_z}{2^{(2L-1)n-k+1}}+\cfrac{1}{2}\right)^2-\cfrac{q_z}{2^{(2L-1)n-k}}+1\right]\\
\label{expect}
&=&\cfrac{1}{2^k}\left[-\cfrac{2}{2^{2[(2L-1)n-k]}}\left(y_z-\cfrac{q_z}{2}\right)^2+\cfrac{3}{2}+\cfrac{2q_z^2}{2^{2[(2L-1)n-k+1]}}\right],
\end{eqnarray}
where in the second equality, we have used $q_z=\sum_{s\in\{0,1\}^{(2L-1)n-k}}g(z,s)=\sum_{s\in\{0,1\}^{(2L-1)n-k}}{\rm Re}[g(z,s)]$.
Therefore, the expected reward is uniquely maximized when
\begin{eqnarray*}
y_z=\cfrac{q_z}{2}.
\end{eqnarray*}

The {\sf BQP} server cannot derive the exact value of $q_z$, which is {\sf $\#$P}-hard in the worst case~\cite{FGHP99}.
However, the {\sf BQP} server can efficiently estimate $q_z$ with polynomial accuracy.
More precisely, for all $z$, the {\sf BQP} server can efficiently obtain $\eta_z$ such that
\begin{eqnarray*}
{\rm Pr}\left[|\eta_z-q_z|\ge\epsilon'\right]\le2e^{-2T{\epsilon'}^2}
\end{eqnarray*}
by sampling from $\{q_z\}_{z\in\{0,1\}^k}$ $T$ times
(for the completeness of the paper, we give the concrete estimation method for $q_z$ in Appendix C)\footnote{It is unknown whether this estimation method is optimal among all methods that can be performed in quantum polynomial time. However, we can say that an optimal method works at least as well as this estimation method. This is sufficient for our purpose.}.
Therefore, $|\eta_z-q_z|\le\epsilon'$ for all $z$ with probability of at least
\begin{eqnarray*}
\left(1-2e^{-2T{\epsilon'}^2}\right)^{2^k}\ge 1-2^{k+1}e^{-2T{\epsilon'}^2}\ge 1-e^{k+1-2T{\epsilon'}^2}.
\end{eqnarray*}
If we set $\epsilon'=1/\{[(2^k+1)f(n)+1]2^k\}$ and $T=(k+1+h(n))/(2{\epsilon'}^2)$, the total repetition number $2^kT$ becomes a polynomial in $n$, and the lower-bound on the probability becomes
\begin{eqnarray*}
1-e^{k+1-2T{\epsilon'}^2}=1-e^{-h(n)}.
\end{eqnarray*}
Therefore, the rational server sends $y_z=\eta_z/2$ to maximize the expected value of the reward $R(y_z,b_z)$ as much as possible.

Finally, we show that when $y_z=\eta_z/2$ and $|\eta_z-q_z|\le\epsilon'$ for all $z$, $\sum_{z\in\{0,1\}^k}|p_z-q_z|\le1/f(n)$.
From Eq.~(\ref{step6}),
\begin{eqnarray*}
p_z=\cfrac{\eta_z}{\sum_{z\in\{0,1\}^k}\eta_z}\le\cfrac{q_z+\epsilon'}{1-2^k\epsilon'}=q_z+\cfrac{q_z2^k\epsilon'+\epsilon'}{1-2^k\epsilon'}
\end{eqnarray*}
and
\begin{eqnarray*}
p_z\ge\cfrac{q_z-\epsilon'}{1+2^k\epsilon'}=q_z-\cfrac{q_z2^k\epsilon'+\epsilon'}{1+2^k\epsilon'}.
\end{eqnarray*}
Therefore,
\begin{eqnarray*}
\label{appro}
|p_z-q_z|\le \cfrac{q_z2^k\epsilon'+\epsilon'}{1-2^k\epsilon'}\le\cfrac{(2^k+1)\epsilon'}{1-2^k\epsilon'}=\cfrac{1}{2^kf(n)}.
\end{eqnarray*}
In conclusion, $\sum_{z\in\{0,1\}^k}|p_z-q_z|\le1/f(n)$.
\hspace{\fill}$\blacksquare$
\\
From Theorem~\ref{sample}, by approximately sampling from $\{p_z\}_{z\in\{0,1\}^k}$, the client can approximately sample from $\{q_z\}_{z\in\{0,1\}^k}$ with high probability.
Given the values of $\{p_z\}_{z\in\{0,1\}^k}$, the approximate sampling from $\{p_z\}_{z\in\{0,1\}^k}$ can be classically performed in polynomial time as shown in Appendix A.
Therefore, as an application of Protocol 1, the classical client can efficiently sample from probability distributions generated by quantum circuits.

In the above proof, we assume that $(1+{\rm Re}[g(z,s)])/2$ can be exactly represented using a polynomial number of bits. If this is not the case, the classical client has to approximate $(1+{\rm Re}[g(z,s)])/2$. As a result, as shown in Appendix B, the expected value of the reward is maximized when $y_z=q_z/2+\delta$, where the real number $\delta$ satisfies $|\delta|\le2^{-f'(n)}$ for a polynomial $f'(n)$.
Therefore, even in the approximation case, the classical client can efficiently obtain the estimated values of the output probabilities of quantum circuits.

Next, we show the following theorem:
\begin{theorem}
Let $L$, $n$, and $k$ be the number of elementary gates used in the quantum circuit $U$ to be executed by the server, the number of input qubits, and the number of measured qubits, respectively.
Let $R(y_z,b_z)$ be the reward function defined in Protocol 1 for $z\in\{0,1\}^k$, $b_z\in\{0,1\}$, and any real value $y_z\in[0,1/2]$.
In Protocol 1, the total reward $\sum_{z\in\{0,1\}^k}R(y_z,b_z)$ is between $3/2-O(1/2^{(2L-1)n-k})$ and $3/2+O(1/2^{(2L-1)n-k})$. Furthermore, the maximum expected value of the total reward is lower-bounded by $3/2+O\left(1/2^{2(2L-1)n-k}\right)$.
\end{theorem}
{\it Proof.} Let $Y_z=(y_z+2^{(2L-1)n-(k+1)})/2^{(2L-1)n-k}$. When $b_z=1$ and $b_z=0$, the total rewards $\sum_{z\in\{0,1\}^k}R(y_z,b_z)$ are $1/2^k\sum_{z\in\{0,1\}^k}2Y_z(2-Y_z)$ and $1/2^k\sum_{z\in\{0,1\}^k}2(1-Y_z^2)$, respectively. Since the client considers $y_z$ to be half of the acceptance probability, we can assume that $0\le y_z\le 1/2$. Therefore, the server should set $Y_z$ in the range from $1/2$ to $1/2+1/2^{(2L-1)n-k+1}$. In this range with $n\ge 1$ and $L\ge 1$, the total reward $\sum_{z\in\{0,1\}^k}R(y_z,b_z)$ is between $3/2-O(1/2^{(2L-1)n-k})$ and $3/2+O(1/2^{(2L-1)n-k})$.

Here, we again assume that $(1+{\rm Re}[g(z,s)])/2$ can be exactly represented using a polynomial number of bits. Note that even if this is not the case, a similar argument holds as shown in Appendix B.
From Eq.~(\ref{expect}), the maximum expected value of the total reward is
\begin{eqnarray*}
\cfrac{1}{2^k}\sum_{z\in\{0,1\}^k}\left(\cfrac{3}{2}+\cfrac{2q_z^2}{2^{2[(2L-1)n-k+1]}}\right)
\ge\cfrac{3}{2}+O\left(\cfrac{1}{2^{2(2L-1)n-k}}\right).
\end{eqnarray*}
Note that even when $y_z$ is an estimated value of $q_z/2$, the expected value of the total reward is lower-bounded by a constant because $\sum_{z\in\{0,1\}^k}R(y_z,b_z)\ge 3/2-O(1/2^{(2L-1)n-k})$.
\hspace{\fill}$\blacksquare$\\
\noindent
From this theorem, Protocol 1 satisfies conditions 1--3 in Sec.~\ref{I}.

\subsection{Decision problems in {\sf BQP}}
\label{IIB}
In this subsection, by applying Protocol 1 in Sec.~\ref{IIA}, we propose a rational delegated quantum computing protocol for decision problems in {\sf BQP}.
To this end, we consider the delegation of the Q-CIRCUIT problem~\cite{ABEM17}.
Since the Q-CIRCUIT problem is a {\sf BQP}-complete problem, any decision problem in {\sf BQP} can be reduced to the Q-CIRCUIT problem.

We set $k=1$ in Protocol 1 in Sec.~\ref{IIA}. Then, by performing Protocol 1 only for $z=1$, the classical client can obtain $\eta$ such that $|\eta-\langle 0^n|U^\dag(|1\rangle\langle 1|\otimes I^{\otimes n-1})U|0^n\rangle|\le1/f(n)$ with probability $1-e^{-h(n)}$ for any polynomials $f(n)$ and $h(n)$.
If $\eta\ge 2/3-1/f(n)$, the client answers YES; if $\eta\le 1/3+1/f(n)$, the client answers NO. Otherwise, the client answers YES or NO uniformly at random. This procedure works because the gap between $2/3-1/f(n)$ and $1/3+1/f(n)$ is at least some constant for sufficiently large $f(n)$.
The client mistakenly answers only when $\eta$ does not satisfy $|\eta-\langle 0^n|U^\dag(|1\rangle\langle 1|\otimes I^{\otimes n-1})U|0^n\rangle|\le1/f(n)$.
Therefore, the probability of the client getting a wrong answer is at most $e^{-h(n)}$.
This means that the classical client can efficiently solve the Q-CIRCUIT problem with the help of the rational quantum server.

\subsection{Estimating output probabilities of approximately sparse quantum circuits with a polynomial number of output qubits}
\label{IIC}
In Sec.~\ref{IIA}, we considered the output probability estimation for any $n$-qubit polynomial-size quantum circuit $U$ with $O(\log{n})$ output qubits.
In this subsection, we consider the same task for a restricted class of $U$ with $n$ output qubits.
More formally, we consider $\epsilon$-approximately $t$-sparse polynomial-size quantum circuits defined in Sec.~\ref{QC}.
Simply speaking, they are $n$-qubit polynomial-size quantum circuits $U$, where there exists a $t$-sparse vector ${\bm v}=(v_z : z\in\{0,1\}^n)$ such that $\sum_{z\in\{0,1\}^n}|q_z-v_z|\le\epsilon$ for output probabilities $q_z\equiv|\langle z|U|0^n\rangle|^2$.

To construct our rational delegated quantum estimating protocol, we show the following theorem{\red:}
\begin{theorem}
\label{list}
Let $\delta=2^{-f(n)}$ for any positive polynomial function $f(n)$ in $n$.
For an $\epsilon$-approximately $t$-sparse polynomial-size $n$-qubit quantum circuit, there exists a polynomial-time quantum algorithm that always outputs a list $\mathcal{L}\equiv\{z^{(1)},\ldots,z^{(l)}\}$, where $l=\lfloor 2t/\epsilon\rfloor$ and each $z^{(i)}$ $(1\le i\le l)$ is an $n$-bit string, such that all $n$-bit strings $z$ satisfying $q_z\ge \epsilon/t$ belong to the list $\mathcal{L}$ with probability of at least $1-\epsilon\delta/(2t+\epsilon)$.
Here, $\lfloor\cdot\rfloor$ is the floor function.
\end{theorem}
{\red Before we show this theorem, we mention a difference between our list $\mathcal{L}$ (in Theorem~\ref{list}) and that obtained in Theorem 10 in Ref.~\cite{SN13}. The list in Ref.~\cite{SN13} satisfies the property that for all elements $z$ in the list, $q_z\ge \epsilon/(2t)$ holds, which is not necessary for our purpose.
Due to this property, it is difficult to efficiently check whether a given list is correct without failing.
By slightly modifying their construction, we circumvent this difficulty and construct a fixed-size list.
The fixed size is necessary to construct our rational delegated quantum estimating protocol.
More precisely, it is used to show Theorem~\ref{replace}.}

{\it Proof.}
Using the method in Theorem 10 in Ref.~\cite{SN13}, the quantum server can efficiently obtain the list $\mathcal{L}'$ with $|\mathcal{L}'|\le\lfloor 2t/\epsilon\rfloor$ such that with probability of at least $1-\epsilon\delta/(2t+\epsilon)$, every $n$-bit string $z$ satisfying $q_z\ge\epsilon/t$ belongs to the list $\mathcal{L}'$.
If $|\mathcal{L}'|=\lfloor 2t/\epsilon\rfloor$, the server sets $\mathcal{L}'=\mathcal{L}$.
On the other hand, if $|\mathcal{L}'|<\lfloor 2t/\epsilon\rfloor$, the server selects $(\lfloor 2t/\epsilon\rfloor-|\mathcal{L}'|)$ $n$-bit strings from $\{z\}_{z\notin \mathcal{L}'}$ in an arbitrary way and incorporates them into $\mathcal{L}'$ to define the set $\mathcal{L}$.
\hspace{\fill}$\blacksquare$

We construct a rational protocol that forces the server to send estimated values $\{\eta_z\}_{z\in \mathcal{L}}$ of $\{q_z\}_{z\in \mathcal{L}}$.
Note that since $t$ and $1/\epsilon$ are polynomials of $n$, the size $|\mathcal{L}|$ of the list is bounded by a polynomial. Therefore, the estimated values $\{\eta_z\}_{z\in \mathcal{L}}$ can be represented using at most a polynomial number of bits.
Furthermore, by using the list $\mathcal{L}$, the estimated values $\{\eta_z\}_{z\in \mathcal{L}}$ can be obtained in quantum polynomial time. 
This is straightforward from Appendix C.
The rational protocol can be constructed by modifying Protocol 1 in Sec.~\ref{IIA} as follows:
\begin{enumerate}
\item[1'.] The rational server selects $\lfloor2t/\epsilon\rfloor$ $n$-bit strings {\red and sends them to the client}. Let $\tilde{\mathcal{L}}$ be the set of the $\lfloor2t/\epsilon\rfloor$ $n$-bit strings. For all $z\in \tilde{\mathcal{L}}$, the rational server and the client perform the following steps:
\begin{enumerate}
\item The rational server sends to the client a non-negative real number $y_z$ that is equal to $\eta_z/2$.
\item The client samples $s$ uniformly at random from $\{0,1\}^{2(L-1)n}$.
\item The client flips a coin that lands heads with probability $(1+{\rm Re}[g(z,s)])/2$. If the coin lands heads, the client sets $b_z=1$; otherwise, $b_z=0$.
\item The client calculates the reward
\begin{eqnarray}
\nonumber
R(y_z,b_z)\equiv\cfrac{1}{\lfloor 2t/\epsilon\rfloor}\Bigg[&&2\cfrac{y_z+2^{2(L-1)n-1}}{2^{2(L-1)n}}b_z+2\left(1-\cfrac{y_z+2^{2(L-1)n-1}}{2^{2(L-1)n}}\right)(1-b_z)\\
\label{reward2}
&&-\left(\cfrac{y_z+2^{2(L-1)n-1}}{2^{2(L-1)n}}\right)^2-\left(1-\cfrac{y_z+2^{2(L-1)n-1}}{2^{2(L-1)n}}\right)^2+1\Bigg].\ \ \ \ \
\end{eqnarray}
Then, the client pays the reward $R(y_z,b_z)$ to the rational server.
\end{enumerate}
\item[2'.] The client calculates
\begin{eqnarray*}
p_z\equiv\cfrac{y_z}{\sum_{z\in\tilde{\mathcal{L}}}y_z}=\cfrac{\eta_z}{\sum_{z\in\tilde{\mathcal{L}}}\eta_z}
\end{eqnarray*}
for all $z\in\tilde{\mathcal{L}}$. On the other hand, for all $z\notin\tilde{\mathcal{L}}$, the client sets $p_z=0$.
\end{enumerate}
\noindent
Since we assume that $t$ and $1/\epsilon$ are polynomials in $n$, the number of repetitions of steps (a)--(d) is bounded by a polynomial.
From Eq.~(\ref{reward2}), in this case, the expected value $\sum_{z\in\tilde{\mathcal{L}}}\mathbb{E}_{b_z}[R(y_z,b_z)]$ of the total reward is
\begin{eqnarray}
\label{rewardsparse}
\cfrac{1}{\lfloor2t/\epsilon\rfloor}\sum_{z\in\tilde{\mathcal{L}}}\left[-2\left(\cfrac{y_z}{2^{2(L-1)n}}-\cfrac{q_z}{2^{2(L-1)n+1}}\right)^2+\cfrac{3}{2}+\cfrac{2q_z^2}{2^{2[2(L-1)n+1]}}\right].
\end{eqnarray}
When $y_z=q_z/2$, Eq.~(\ref{rewardsparse}) is maximized and a monotonically increasing function of $q_z$ in the range of $q_z\ge 0$.
Note that $q_z\ge 0$ holds because it is a probability.
Therefore, to increase the expected value of the total reward, the rational server has to include all bit strings whose probabilities are larger than $\epsilon/t$ into the list $\tilde{\mathcal{L}}$.
Therefore, from Theorem~\ref{list} and Appendix C, the rational quantum server can efficiently generate such list $\tilde{\mathcal{L}}$ and $\{\eta_z\}_{z\in \tilde{\mathcal{L}}}$.

It is worth mentioning that we can obtain the same conclusion as above even when $y_z$ is an estimated value of $q_z/2$, i.e., $|y_z-q_z/2|\le\sqrt{\epsilon'}$ with probability of at least $1-\delta$, where $\epsilon'$ and $\delta$ are $1/n^{O(1)}$ and $1/2^{n^{O(1)}}$, respectively.
When $y_z$ is the estimated value, the server may be able to increase the expected value of the reward by including a bit string whose probability is small.
Consider the following situations: when the true value is $q_z=0.99$, the estimated value can become $y_z=0$ with a non-zero probability. In this case, the expected value $\mathbb{E}_{b_z}[R(y_z,b_z)]$ is $3/2$.
On the other hand, when the true value is $q_z=0.01$, the estimated value can become $y_z=0.01$ with a non-zero probability.
In this case, the expected value is larger than $3/2$.
This example implies that we have to appropriately take the case where the estimation fails into account.

We show the following theorem:
\begin{theorem}
\label{replace}
Let $q_z\equiv|\langle z|U|0^n\rangle|^2$ for $\epsilon$-approximately $t$-sparse polynomial-size quantum circuits $U$ and $z\in\{0,1\}^n$.
Even if $y_z$ is an estimated value of $q_z/2$, all bit strings whose probabilities are larger than $\epsilon/t$ must be included into the list $\tilde{\mathcal{L}}$ to maximize the expectation value of the total reward $\sum_{z\in\tilde{\mathcal{L}}}R(y_z,b_z)$, where $R(y_z,b_z)$ is defined in Eq.~(\ref{reward2}).
\end{theorem}
{\it Proof.}
Let us assume that an outcome $z_1$ with $q_{z_1}\ge\epsilon/t$ is not included in the list $\tilde{\mathcal{L}}$.
The list $\tilde{\mathcal{L}}$ includes at least one $z_2$ such that $q_{z_2}\le1/\lfloor2t/\epsilon\rfloor$ because $|\tilde{\mathcal{L}}|=\lfloor2t/\epsilon\rfloor$.
We show that the server can increase the expected value of the reward by replacing $z_2$ with $z_1$.
This means that the rational server includes all probabilities larger than $\epsilon/t$ into $\tilde{\mathcal{L}}$.

From Eq.~(\ref{reward2}), when $q_z=q_{z_1}$, the expected value $\mathbb{E}[R(y_{z_1},b_{z_1})]$ of the reward is lower-bounded by
\begin{eqnarray*}
\cfrac{1-\delta}{\lfloor2t/\epsilon\rfloor}\left(-\cfrac{2\epsilon'}{2^{2[2(L-1)n+1]}}+\cfrac{3}{2}+\cfrac{2q_{z_1}^2}{2^{2[2(L-1)n+1]}}\right)
+\cfrac{\delta}{\lfloor2t/\epsilon\rfloor}\left(-\cfrac{2}{2^{2[2(L-1)n+1]}}+\cfrac{3}{2}+\cfrac{2q_{z_1}^2}{2^{2[2(L-1)n+1]}}\right),
\end{eqnarray*}
where we take the case where the estimation fails into account.
On the other hand, when $q_z=q_{z_2}$, the expected value $\mathbb{E}[R(y_{z_2},b_{z_2})]$ of the reward is upper-bounded by
\begin{eqnarray*}
\cfrac{1}{\lfloor2t/\epsilon\rfloor}\left(\cfrac{3}{2}+\cfrac{2q_{z_2}^2}{2^{2[2(L-1)n+1]}}\right).
\end{eqnarray*}
Here, we set $\epsilon'<\epsilon^2(3t^2+\epsilon^2-4t\epsilon)/[2t^2(2t-\epsilon)^2]$.
The gap between these two expected values is
\begin{eqnarray*}
\mathbb{E}[R(y_{z_1},b_{z_1})]-\mathbb{E}[R(y_{z_2},b_{z_2})]&\ge&\cfrac{2\left\{(q_{z_1}^2-q_{z_2}^2)-[(1-\delta)\epsilon'+\delta]\right\}}{\lfloor2t/\epsilon\rfloor 2^{2[2(L-1)n+1]}}\\
&\ge&\cfrac{2\left\{\epsilon^2/t^2-[\epsilon/(2t-\epsilon)]^2-[(1-\delta)\epsilon'+\delta]\right\}}{\lfloor2t/\epsilon\rfloor 2^{2[2(L-1)n+1]}}\\
&\ge&\cfrac{2}{\lfloor2t/\epsilon\rfloor 2^{2[2(L-1)n+1]}}\left[\cfrac{\epsilon^2(3t^2+\epsilon^2-4t\epsilon)}{t^2(2t-\epsilon)^2}-2\epsilon'\right]\\
&>&0,
\end{eqnarray*}
where in the third inequality we have used $\epsilon'\ge\delta$.
The above argument holds even in the approximation case discussed in Appendix B.
\hspace{\fill}$\blacksquare$

In the remainder of this section, using the result in Ref.~\cite{SN13}, we show that the client can obtain approximated values of $\{q_z\}_{z\in\{0,1\}^n}$ from $\{\eta_z\}_{z\in \mathcal{L}}$.
To this end, we first show the following theorem:
\begin{theorem}
\label{sparseapp}
Let $t$ and $1/\epsilon\ {\red(\ge 6)}$ be any positive polynomials in $n$ and $\delta=2^{-f(n)}$ for any positive polynomial function $f(n)$ in $n$.
Let $\mathcal{Q}\equiv\{q_z\}_{z\in\{0,1\}^n}$ be the output probability distribution of an $\epsilon$-approximately $t$-sparse polynomial-time quantum circuit with $n$ input qubits.
Given the list $\mathcal{L}$ defined in Theorem~\ref{list}, a $\lfloor 2t/\epsilon\rfloor$-sparse vector $\{\eta_z\}_{z\in\{0,1\}^n}$ that satisfies
{\red\begin{eqnarray*}
\sum_{z\in\{0,1\}^n}|q_z-\eta_z|\le 3\epsilon
\end{eqnarray*}}
with probability at least $1-\delta$ can be obtained in quantum polynomial time.
\end{theorem}

The following theorem also holds:
\begin{theorem}
\label{sparsesamp}
Parameters $\epsilon$, $t$, and $\delta$ are set as in Theorem~\ref{sparseapp}.
Let $\mathcal{Q}\equiv\{q_z\}_{z\in\{0,1\}^n}$ be the output probability distribution of an $\epsilon$-approximately $t$-sparse polynomial-time quantum circuit with $n$ input qubits.
Using the $\lfloor2t/\epsilon\rfloor$-sparse vector $\{\eta_z\}_{z\in\{0,1\}^n}$ obtained in Theorem~\ref{sparseapp}, it is possible to efficiently classically compute an $\lfloor2t/\epsilon\rfloor$-sparse probability distribution $\mathcal{P}\equiv\{p_z\}_{z\in\{0,1\}^n}$ that satisfies
{\red\begin{eqnarray*}
\sum_{z\in\{0,1\}^n}|q_z-p_z|\le 12\epsilon
\end{eqnarray*}}
with probability of at least $1-\delta$.
\end{theorem}
The combination of Theorems~\ref{sparseapp} and \ref{sparsesamp} 
is the same as Theorem 11 in Ref.~\cite{SN13} except for the type of list used.
The proofs of Theorems~\ref{sparseapp} and \ref{sparsesamp} are essentially the same as the proof of Theorem 11 in Ref.~\cite{SN13}. For the completeness of this paper, we give them in Appendices D and E.

From Appendix D, it is known that $\eta_z=0$ for all $z\notin\mathcal{L}$.
Therefore, from Theorem~\ref{sparsesamp}, if $\{\eta_z\}_{z\in\mathcal{L}}$ is given, the approximate sampling from an output probability distribution of an $\epsilon$-approximately $t$-sparse polynomial-size quantum circuit can be performed in classical polynomial time using the method in Appendix A.
Note that although the sampling method in Appendix A is tailored for a probability distribution $\{q_z\}_{z\in\{0,1\}^k}$ on $k=O(\log{n})$ bit strings, it works even if $\{q_z\}_{z\in\{0,1\}^k}$ is replaced with the polynomially sparse probability distribution $\{\eta_z\}_{z\in\{0,1\}^n}$.

The efficient classical simulatability of approximately sparse quantum circuits was explored in Ref.~\cite{SN13}. However, their classical-simulation algorithm requires some additional constraints for quantum circuits.
In general, approximately sparse quantum circuits are not known to be efficiently classically simulatable.

\section{Multi-rational-server delegated quantum computing with a constant reward gap}
\label{IIIC}
In this section, we consider the reward gap.
Although a large reward gap is, in practice, desirable to incentivize the server to behave optimally, our sumcheck-based protocols have only exponentially small gaps.
The existing rational delegated quantum computing protocols~\cite{MN18} also have only exponentially small gaps.
It is open as to whether a constant (or $1/n^{O(1)}$) reward gap is possible.
However, in this subsection, we show that if non-communicating but entangled multiservers are allowed, we can construct a rational delegated quantum computing protocol with a constant reward gap for {\sf BQP} problems while keeping the first three conditions 1--3 in Sec.~\ref{I}.
To this end, we utilize multiprover interactive proof systems for {\sf BQP}.
In some multiprover interactive proof systems proposed for {\sf BQP}, the computational ability of the honest provers is bounded by {\sf BQP} but that of the malicious provers is unbounded~\cite{FHM18,RUV13,GKW15,CGJV17,NV17,NV18,G19}\footnote{In this paper, we focus on multiprover interactive proof systems that consist of a constant number of provers. Some multiprover interactive proof systems~\cite{M16,HPF15} require polynomially many provers.}.
Simply speaking, these multiprover interactive proof systems satisfy the following theorem:
\begin{theorem}[\cite{FHM18,RUV13,M16,GKW15,HPF15,CGJV17,NV17,NV18,G19}]
For any language $L\in${\sf BQP}, there exists a $poly(|x|)$-time classical verifier $V$ interacting with a constant number of non-communicating but entangled provers, such that for inputs $x$,
\begin{enumerate}
\item if $x\in L$, then there exists a $poly(|x|)$-time quantum provers' strategy in which $V$ accepts with probability at least $2/3$
\item if $x\notin L$, then for any (computationally-unbounded) provers' strategy, $V$ accepts with probability at most $1/3$.
\end{enumerate}
Here, $|x|$ denotes the size (i.e., the bit length) of $x$.
\label{MIPM16}
\end{theorem}
We denote the above interaction between $V$ and provers as $\pi_L$ for the language $L\in${\sf BQP}.

Using the above multiprover interactive proof systems, we construct the following rational delegated quantum computing protocol\footnote{This construction is essentially the same as that used in Ref.~\cite{AM12} to show {\sf IP}$\subseteq${\sf RIP}. Here, {\sf RIP} is the complexity class of decision problems that can be solved by rational interactive proof systems whose prover's computational power is unbounded.}:\\
\noindent
[{\bf Protocol 2}]\\
This protocol depends on the input size $|x|$, the number $M$ of servers, and the interaction $\pi_L$.
\begin{enumerate}
\item For a given {\sf BQP} language $L$ and an instance $x$, the first server among $M$ rational ones sends $b\in\{0,1\}$ to the client. As shown in Theorem~\ref{multiconstant}, if the server is rational, $b=1(0)$ when $x$ is in $L$ ($x$ is not in $L$).
\item If $b=1$, the client and $M$ servers simulate $\pi_L$ for the language $L$ and instance $x$; otherwise, the client and $M$ servers simulate $\pi_{\bar{L}}$ for the complement $\bar{L}$ and the instance $x$.
More precisely, the client and servers simulate the verifier and provers in $\pi_L$ or $\pi_{\bar{L}}$, respectively.
\item The client pays reward $R=1/M$ to each of the $M$ servers if the simulated verifier accepts. On the other hand, if the simulated verifier rejects, the client pays $R=0$.
\item The client concludes $x\in L$ if $b=1$; otherwise, the client concludes $x\notin L$.
\end{enumerate}
Note that since {\sf BQP} is closed under complement, $\pi_{\bar{L}}$ exists for the complement $\bar{L}$.
Here, we notice that even if the simulated verifier accepts, each server can obtain only $1/M$ as the reward.
However, since the number $M$ of the servers is two in the multiprover interactive proof systems in Refs.~\cite{RUV13,GKW15,CGJV17,G19}, the reward $1/M$ paid to each server can be made $1/2$.
Particularly, when we use multiprover interactive proof systems in Refs.~\cite{CGJV17,G19} among them, the number of rounds in Protocol 2 becomes a constant.

We clarify the meaning of ``rational" in multi-rational-server delegated quantum computing.
In this computing model, we can consider at least two possible definitions of ``rational."
One is that each server wants to maximize each reward, and
the other is that all servers want to collaboratively maximize their total reward.
Fortunately, in Protocol 2, these two definitions are equivalent.
In other words, the total reward is maximized if and only if the reward paid to each server is maximized.
Hereafter, we therefore do not distinguish between these two definitions.

Before we show that Protocol 2 has a constant reward gap, we show that if the servers are rational, the client's answer is correct.
More formally, we prove the following theorem:
\begin{theorem}
\label{multiconstant}
Let $b\in\{0,1\}$, $L\in${\sf BQP}, and $x$ be an instance of $L$.
In Protocol 2, if the servers are rational, i.e., take the strategy that maximizes the expectation value of the reward, then $b=1$ if and only if $x\in L$.
\end{theorem}
{\it Proof.}
Before the formal proof, we explain our proof idea.
From Protocol 2, we can know that the reward is paid to the servers only when the simulated verifier accepts.
This implies that the servers would like to select an interaction from $\pi_L$ and $\pi_{\bar{L}}$ such that the verifier accepts with higher probability.
From Theorem~\ref{MIPM16}, when $x\in L$ ($x\notin L$), the verifier in $\pi_L$ ($\pi_{\bar L}$) accepts with higher probability than that in the other.
Therefore, when $x\in L$ ($x\notin L$), the first server sends $b=1$ ($b=0$) to the client.

The formal proof is as follows.
First, we consider the YES case, i.e., the case where $x$ is in $L$. 
If $b=1$, the client and the servers perform $\pi_L$ for the language $L$ and the instance $x$. Therefore, when the servers simulate the honest provers in $\pi_L$, the client accepts with probability of at least $2/3$.
On the other hand, if $b=0$, the client accepts with probability less than or equal to $1/3$.
This is because $x$ is a NO instance for the complement $\bar{L}$, i.e., $x\notin\bar{L}$.
In $\pi_{\bar{L}}$, when the answer is NO, the acceptance probability is at most $1/3$ for any provers' strategy.
Since the completeness-soundness gap $1/3$ is a positive constant, one of the rational servers sends $b=1$ if $x\in L$.
By following the same argument, one of them sends $b=0$ when $x\notin L$.
\hspace{\fill}$\blacksquare$

From this proof, we notice that the reward gap has the same value as the completeness-soundness gap\footnote{Precisely speaking, since the computational power of the server is bounded by {\sf BQP}, the server sends $b=0(1)$ with an exponentially small probability when the correct answer is YES (NO). Therefore, the finally obtained reward gap is decreased by $1/2^{n^{O(1)}}$ from the original completeness-soundness gap. However, this gap is negligible because the original completeness-soundness gap is a constant. Note that {\red we say a function $f(x)$} negligible if ${\red |f(x)|}< 1/p(x)$ holds for any polynomial function $p(x)$ and all sufficiently large $x$.}.
Protocol 2 has a $1/3$ reward gap, which is constant.
Furthermore, it can be straightforwardly shown that Protocol 2 also satisfies conditions 1--3 mentioned in Sec.~\ref{I} as follows.
Since the total reward $M\times R$ paid to $M$ servers is $0$ or $1$, the first and second conditions are satisfied.
When the servers behave rationally, the client accepts with probability at least $2/3$.
Therefore, the expected value of the total reward paid to the rational servers is at least $2/3$, which satisfies the third condition.

\section{Relation between rational and ordinary delegated quantum computing protocols}
\label{X}
In Sec.~\ref{IIIC}, by incorporating {\it ordinary} delegated quantum computing into {\it rational} delegated quantum computing, we have shown that the four conditions can be simultaneously satisfied.
In this section, we consider the reverse direction, i.e., constructing {\it ordinary} delegated quantum computing protocols from {\it rational} delegated quantum computing protocols.
By combining this construction with the result in Sec.~\ref{IIIC}, we obtain an equivalence (under a certain condition) between these two types of delegated quantum computing.
Note that in ordinary ones, the server's ability is unbounded in NO cases (i.e., when $x\notin L$).

To construct ordinary delegated quantum computing protocols from rational ones, we consider the general $poly(|x|)$-round rational delegated quantum computing protocol defined in Definition~\ref{RDQCRDQC}, which we call RDQC for short.
By applying two additional conditions for RDQC, we define constrained RDQC as follows:
\begin{definition}
\label{constrained}
Let $c_{\rm YES}$ be a positive constant defined in Eq.~(\ref{cyes}).
Let $s$, $S_{\rm incorrect}$, and $R(s,x)$ be the server's strategy, the set of the server's strategies that make the client output an incorrect answer with unit probability, and a reward function depending on $s$ and an instance $x$ of $L$, respectively.
For $L\in${\sf BQP}, the constrained RDQC protocol is an RDQC protocol defined in Definition~\ref{RDQCRDQC} such that
\begin{enumerate}
\item There exists a classically efficiently computable positive polynomial $f(|x|)$ such that
\begin{eqnarray}
\label{add}
c_{\rm YES}-{\rm max}_{s\in S_{\rm incorrect},x\notin L}\mathbb{E}[R(s,x)]\ge\cfrac{1}{f(|x|)},
\end{eqnarray}
\item The upper-bound $c$ of the reward is classically efficiently computable.
\end{enumerate}
\end{definition}
The first condition was introduced in Ref.~\cite{CMS18}.
It is worth mentioning that the second condition is satisfied in our sumcheck-based rational protocols, while the first condition is not.
Note that the left-hand side of Eq.~(\ref{add}) is not the reward gap.

We show that an {\it ordinary} delegated quantum computing protocol with a single {\sf BQP} server and a single {\sf BPP} client can be constructed from any constrained RDQC protocol.
This means that if we can construct a constrained RDQC protocol, then we can affirmatively solve the open problem of whether a classical client can efficiently delegate universal quantum computing to a quantum server while efficiently verifying his/her integrity.
To this end, we show the following theorem:
\begin{theorem}
\label{nogo}
The function $f(|x|)$ and the parameter $c$ are set as in Definition~\ref{constrained}.
If a language $L$ in {\sf BQP} has a $k$-round constrained RDQC protocol, then $L$ has a $k$-round interactive proof system with the completeness-soundness gap $1/(cf(|x|))$ between an honest {\sf BQP} prover and a {\sf BPP} verifier.
\end{theorem}
The proof is essentially the same as that of Theorem 4 in Ref.~\cite{CMS18}.
We defer it to Appendix F.

As a corollary from Theorem~\ref{nogo}, it seems that a constant-round constrained RDQC protocol cannot exist for {\sf BQP}.
More concretely, we obtain the following corollary:
\begin{corollary}
\label{constantng}
If there exists a constant-round constrained RDQC protocol for {\sf BQP}, then {\sf BQP}$\subseteq{\sf \prod_2^p}$.
\end{corollary}
{\it Proof.}
We use essentially the same argument as in Ref.~\cite{FHM18}.
From Theorem~\ref{nogo}, if there exists a $k$-round constrained RDQC protocol for {\sf BQP}, then {\sf BQP} is contained in {\sf IP}$[k]$, which is a class of decision problems having a $k$-round interactive proof system.
From known results~\cite{GS86,B85,BM88}, {\sf IP}$[k]\subseteq${\sf AM}$[k+2]=${\sf AM}$[2]\subseteq{\sf \prod_2^p}$ with a constant $k$.
\hspace{\fill}$\blacksquare$
\\Given the oracle separation between {\sf BQP} and {\sf PH}~\cite{RT19}, the inclusion {\sf BQP}$\subseteq{\sf \prod_2^p}$ is considered unlikely.

In Theorem~\ref{nogo}, we show that a constrained RDQC protocol can be converted to an {\it ordinary} delegated quantum computing protocol.
We show that the reverse conversion is also possible using the idea in Sec.~\ref{IIIC}.
\begin{theorem}
\label{gyaku}
If a language $L$ in {\sf BQP} has an interactive proof system with an honest {\sf BQP} prover and a {\sf BPP} verifier, then $L$ has a constrained RDQC protocol whose reward gap is exponentially close to the completeness-soundness gap of the original interactive proof system.
\end{theorem}
{\it Proof.} 
First, we explain our proof idea for clarity.
From the assumption, we have an interactive proof system with an honest {\sf BQP} prover and a {\sf BPP} verifier.
By replacing the multiprover interactive proof system in the argument of Sec.~\ref{IIIC} with this interactive proof system, we can construct a RDQC protocol.
The remaining task is to show that it satisfies Eq.~(\ref{add}).
This construction converts the completeness-soundness gap in the original protocol into the gap in Eq.~(\ref{add}) with almost no changes in the size of the gap.
It implies that the resultant gap is larger than the inverse of any positive polynomial in $|x|$.
 
The formal proof is as follows:
Let assume that any language $L$ in {\sf BQP} has an interactive proof system with an honest {\sf BQP} prover and a {\sf BPP} verifier.
In other words, for any language $L\in${\sf BQP}, there exists a polynomial-time classical verifier interacting with a prover, such that for inputs $x$, if $x\in L$, then there exists a {\sf BQP} prover's strategy, where the verifier accepts with probability of at least $c'$, and if $x\notin L$, then for any computationally-unbounded prover's strategy, the verifier accepts with probability of at most $s'$. Here, $c'-s'$ is at least a constant.
We denote this interaction between the prover and the verifier as $\tilde{\pi}_L$ for the language $L\in${\sf BQP}.
We simply replace the multiprover interactive proof system in the argument of Sec.~\ref{IIIC} with the interactive proof system for {\sf BQP}.
As a result, we obtain the following RDQC protocol for {\sf BQP}:
\begin{enumerate}
\item For a given {\sf BQP} language $L$ and an instance $x$, the rational server sends $b\in\{0,1\}$ to the client.
\item If $b=1$, the client and server simulate $\tilde{\pi}_L$ for the language $L$ and instance $x$; otherwise, they simulate $\tilde{\pi}_{\bar{L}}$ for the complement $\bar{L}$ and the instance $x$.
\item The client pays reward $R=1$ to the server if the simulated verifier accepts. On the other hand, if the simulated verifier rejects, the client pays $R=0$.
\item The client concludes $x\in L$ if $b=1$; otherwise, the client concludes $x\notin L$.
\end{enumerate}
Note that since {\sf BQP} is closed under complement, $\tilde{\pi}_{\bar{L}}$ exists for the complement $\bar{L}$.

Since the upper-bound of the reward is obviously one, the remaining task is to show that the constructed RDQC protocol satisfies Eq.~(\ref{add}).
Let $c'$ and $s'$ be the completeness and soundness parameters of the interactive proof system for {\sf BQP}, respectively.
From the construction, if the server is rational, $b=1(0)$ with probability exponentially close to one when $x\in L$ ($x\notin L$).
Therefore, $c_{\rm YES}$ and ${\rm max}_{s\in S_{\rm incorrect},x\notin L}\mathbb{E}[R(s,x)]$ are identical with $c'(1-o(1))$ and $s'$, respectively.
As a result, $c_{\rm YES}-{\rm max}_{s\in S_{\rm incorrect},x\notin L}\mathbb{E}[R(s,x)]=c'(1-o(1))-s'>1/{\rm poly}(|x|)$.

By using the similar argument, we can also show that the reward gap is exponentially close to $c'-s'$.
For any $x$, the rational server sends $b=1(0)$ with probability exponentially close to one when $x\in L$ ($x\notin L$).
Therefore, the maximized expected value of the reward is at least exponentially close to $c'$.
On the other hand, to make the client output an incorrect answer, the server has to send $b=0(1)$ when $x\in L$ ($x\notin L$).
In this case, the expected value of the reward is at most $s'$.
From the above argument, the reward gap is exponentially close to $c'-s'$.
\hspace{\fill}$\blacksquare$\\
\noindent
From Theorems~\ref{nogo} and \ref{gyaku}, constrained RDQC and ordinary delegated quantum computing with an honest {\sf BQP} prover and a {\sf BPP} verifier are convertible from one to the other and vice versa.
As an interesting observation from a reviewer of this paper, these conversions can be done even if Eq.~(\ref{cno}) is removed from Definition~\ref{RDQCRDQC}.

Finally, by applying Theorems~\ref{nogo} and \ref{gyaku}, we give the following amplification method for the reward gap:
\begin{corollary}
\label{RGamp}
The reward gap of the constrained RDQC can be efficiently amplified to a constant from any value.
\end{corollary}
{\it Proof.}
First, we convert a constrained RDQC protocol, whose reward gap is at most some constant, to an interactive proof system for {\sf BQP} using Theorem~\ref{nogo}.
Then using the standard amplification method for the completeness-soundness gap (i.e., the repetition and the majority vote)~\cite{GMR89}, we obtain the interactive proof system with a constant completeness-soundness gap.
Finally, using Theorem~\ref{gyaku}, we convert it to another constrained RDQC protocol.
From the conversion method used in the proof of Theorem~\ref{gyaku}, it is obvious that conditions 1--3 in Sec.~\ref{I} and conditions 1--2 in Definition~\ref{constrained} are satisfied.
Since the finally obtained reward gap is exponentially close to the completeness-soundness gap in this conversion, the finally obtained constrained RDQC protocol has a constant reward gap.
\hspace{\fill}$\blacksquare$\\
\noindent
As an interesting point, this amplification method works even if the original constrained RDQC protocol has only an exponentially small reward gap.
This is because the original constrained RDQC protocol satisfies Eq.~(\ref{add}).
Note that since the finally obtained constrained RDQC protocol is no longer a one-round one, Corollary~\ref{RGamp} circumvents the no-go result in Ref.~\cite{MN18}.

\section{Conclusion}
\label{dis}
We conclude this paper by discussing another way of achieving a constant reward gap and presenting our outlook.
\subsection{Discussion}
\label{IIIB}
An idea similar to that in Sec.~\ref{IIIC} can be used to construct a single-server rational delegated quantum computing protocol with a constant reward gap for {\sf BQP} if we assume that the LWE problem is hard for polynomial-time quantum computing.
Note that the hardness of the LWE problem is widely accepted in the fields of quantum cryptography~\cite{M18,CCKW19,GV19} and modern cryptography~\cite{R09}.
To this end, we utilize Mahadev's result in Ref.~\cite{M18}.
Recently, for all {\sf BQP} problems, Mahadev has constructed an interactive argument system with a constant completeness-soundness gap under the hardness assumption on LWE problems.
In other words, for any language $L\in${\sf BQP}, there exists a polynomial-time classical verifier interacting with a polynomial-time quantum prover, such that for inputs $x$, if $x\in L$, then there exists a {\sf BQP} prover's strategy, where the verifier accepts with probability of at least $c'$, and if $x\notin L$, then for any {\sf BQP} prover's strategy, the verifier accepts with probability of at most $s'$. Here, $c'-s'$ is at least a constant.
We denote this interaction between the prover and verifier as $\pi'_L$ for the language $L\in${\sf BQP}.
In Mahadev's interactive argument system, the prover's computational ability is bounded by polynomial-time quantum computing regardless of whether $x\in L$ or $x\notin L$.

To construct a single-server rational delegated quantum computing protocol with a constant reward gap for {\sf BQP}, we simply replace the multiprover interactive proof system in the argument in Sec.~\ref{IIIC} with Mahadev's interactive argument system as follows:\\
\noindent
[{\bf Protocol 3}]\\
This protocol depends on the instance size $|x|$ and the interaction $\pi'_L$.
\begin{enumerate}
\item For a given {\sf BQP} language $L$ and an instance $x$, the rational server sends $b\in\{0,1\}$ to the client.
\item If $b=1$, the client and server simulate $\pi'_L$ for the language $L$ and instance $x$; otherwise, they simulate $\pi'_{\bar{L}}$ for the complement $\bar{L}$ and the instance $x$.
\item The client pays reward $R=1$ to the server if the simulated verifier accepts. On other hand, if the simulated verifier rejects, the client pays $R=0$.
\item The client concludes $x\in L$ if $b=1$; otherwise, the client concludes $x\notin L$.
\end{enumerate}
Note that since {\sf BQP} is closed under complement, $\pi'_{\bar{L}}$ exists for the complement $\bar{L}$.
Furthermore, since Mahadev's interactive argument system is a constant-round protocol, Protocol 3 is also a constant-round one.

For clarity, we remark that the above constructed rational delegated quantum computing protocol does not work if the server's computational ability is unbounded.
This comes from the definition of the interactive argument system--if the malicious prover's computational ability is unbounded, the malicious prover may be able to make the verifier conclude YES with a high probability when the correct answer is NO.

From the proof of Theorem~\ref{multiconstant}, we know that the reward gap is exponentially close to the completeness-soundness gap.
Since Mahadev's interactive argument system has a constant completeness-soundness gap, the reward gap of Protocol 3 is constant.
Furthermore, since the completeness parameter $c'$ is negligibly close to one in Mahadev's interactive argument system, Protocol 3 also satisfies conditions 1--3 mentioned in Sec.~\ref{I}.

We here again note that in the LWE-based rational delegated quantum computing protocol, elements in the set $S_{\rm incorrect}$ are restricted to strategies that can be performed by a polynomial-time quantum server. On the other hand, in other rational protocols presented in this paper, such a restriction is not necessary.

\subsection{Outlook}
In this paper, we have considered the integrity of cloud quantum computing.
Another important notion in cloud quantum computing is blindness, which means to delegate quantum computing to a remote server while hiding inputs, outputs, and quantum algorithms.
When we require information-theoretic security for cloud quantum computing, classical computing seems not to be sufficient for the client~\cite{MK19,ACGK17,MNTT18}.
On the other hand, if we assume that LWE problems are difficult for efficient quantum computing, the classical client can perform verifiable blind quantum computing that is secure against a polynomial-time quantum adversary~\cite{GV19,CCKW19}. Although the rational protocols proposed in this paper are not blind, it would be interesting to consider whether a classical-client verifiable blind quantum computing protocol can be constructed when we assume that a server is rational.

\section{Acknowledgments}
We thank Yupan Liu, Seiseki Akibue, Harumichi Nishimura, and Fran\c{c}ois Le Gall for helpful discussions. We also thank Yasuhiro Takahashi for fruitful discussions and pointing out Ref.~\cite{SN13} to us.
This work is supported by JST [Moonshot R\&D--MILLENNIA Program] Grant Number JPMJMS2061.
YT is supported by MEXT Quantum Leap Flagship Program (MEXT Q-LEAP) Grant Number JPMXS0118067394 and JPMXS0120319794. TM is supported by MEXT Quantum Leap Flagship Program (MEXT Q-LEAP) Grant Number JPMXS0118067394 and JPMXS0120319794, JST PRESTO No.JPMJPR176A, JSPS Grant-in-Aid for Young Scientists (B) No.JP17K12637, and the Grant-in-Aid for Scientific Research (B) No.JP19H04066 of JSPS.
ST is supported by JSPS KAKENHI Grant Number JP20H05966.

\section{Appendix A: Approximate sampling from a known probability distribution}
In this Appendix, we give an efficient method to approximately sample from a known probability distribution $\{t_s\}_{s\in\{0,1\}^k}$, where $k$ is at most $O(\log{n})$.
In step (c) of Protocol 1 in Sec.~\ref{IIA}, $k=1$ and $t_0=(1+{\rm Re}[g(z,s)])/2$.
If the client performs the approximate sampling from $\{p_z\}_{z\in\{0,1\}^k}$ in Eq.~(\ref{step6}), $k=O(\log{n})$ and $t_s=p_s$.
The following algorithm approximately samples from $\{t_s\}_{s\in\{0,1\}^k}$ in classical polynomial time:
\begin{enumerate}
\item Approximate each $t_s$ using $m(\ge 2k)$ bits as follows.
\begin{enumerate}
\item Find a single $s_{\rm max}$ such that $t_{s_{\rm max}}\ge t_s$ for any $s\neq s_{\rm max}$.
\item For all $s$ except for $s_{\rm max}$, the probability $t_s$ is approximated as the form
\begin{eqnarray}
\label{binary}
\tilde{t}_s\equiv\sum_{j=1}^m2^{-j}a_j^{(s)}
\end{eqnarray}
that satisfies $|\tilde{t}_s-t_s|\le2^{-m}$, and $a_j^{(s)}\in\{0,1\}$.
For $s_{\rm max}$, $\tilde{t}_{s_{\rm max}}\equiv 1-\sum_{s\neq s_{\rm max}}\tilde{t}_s$, which can also be represented as the form of Eq.~(\ref{binary}).
\end{enumerate}
\item Uniformly and randomly generate an $m$-bit string $w_1\ldots w_m\in\{0,1\}^m$.
\item Output $s$ such that
\begin{eqnarray*}
\sum_{y<s}\tilde{t}_y\le\sum_{j=1}^m2^{-j}w_j<\sum_{y\le s}\tilde{t}_y,
\end{eqnarray*}
where $y<s$ and $y\le s$ if $\sum_{j=1}^m2^jy_j<\sum_{j=1}^m2^js_j$ and $\sum_{j=1}^m2^jy_j\le\sum_{j=1}^m2^js_j$, respectively.
Here, $y_j$ and $s_j$ are the $j$th bits of the $m$-bit strings $y$ and $s$, respectively.
\end{enumerate}

To show that this algorithm approximately samples from $\{t_s\}_{s\in\{0,1\}^k}$, we derive an upper-bound of $\sum_{s\in\{0,1\}^k}|t_s-\tilde{t}_s|$.
From $|\tilde{t}_s-t_s|\le2^{-m}$ for all $s$ except for $s_{\rm max}$,
\begin{eqnarray*}
t_s-2^{-m}\le& \tilde{t}_s&\le t_s+2^{-m}\\
\sum_{s\neq s_{\rm max}}(t_s-2^{-m})\le& \sum_{s\neq s_{\rm max}}\tilde{t}_s&\le \sum_{s\neq s_{\rm max}}(t_s+2^{-m})\\
(1-t_{s_{\rm max}})-(2^k-1)2^{-m}\le& 1-\tilde{t}_{s_{\rm max}}&\le (1-t_{s_{\rm max}})+(2^k-1)2^{-m}\\
t_{s_{\rm max}}-(2^k-1)2^{-m}\le& \tilde{t}_{s_{\rm max}}&\le t_{s_{\rm max}}+(2^k-1)2^{-m}.
\end{eqnarray*}
Therefore, $\sum_{s\in\{0,1\}^k}|t_s-\tilde{t}_s|\le (2^k-1)2^{1-m}$.
This means that for polynomially increasing $m$, the algorithm can
sample from $\{t_s\}_{s\in\{0,1\}^k}$ with exponential precision.
Here, we note that $\tilde{t}_{s_{\rm max}}$ is not negative because $t_{s_{\rm max}}\ge 1/2^k$ and $m\ge 2k$.

\section{Appendix B: Approximation case}
In this Appendix, we consider the case where $(1+{\rm Re}[g(z,s)])/2$ cannot be exactly represented using a polynomial number of bits. In this case, using the method in Appendix A, the classical client can sample from $\{\tilde{t}_0,\tilde{t}_1\}$ with $\tilde{t}_0-(1+{\rm Re}[g(z,s)])/2=\delta$ and $\tilde{t}_1=1-\tilde{t}_0$. Here, the real number $\delta$ can be set to satisfy $|\delta|\le 2^{-f'(n)-(2L-1)n+k}$ with any polynomial $f'(n)$. Therefore,
\begin{eqnarray}
\nonumber
&&2^k\times\mathbb{E}_{b_z}[R(y_z,b_z)]\\
\nonumber
&=&\sum_{s\in\{0,1\}^{(2L-1)n-k}}\cfrac{1}{2^{(2L-1)n-k}}\Bigg\{\left(\cfrac{1+{\rm Re}[g(z,s)]}{2}+\delta\right)\left[2Y_z-Y_z^2-\left(1-Y_z\right)^2+1\right]\\
\nonumber
&&+\left(\cfrac{1-{\rm Re}[g(z,s)]}{2}-\delta\right)\left[2\left(1-Y_z\right)-Y_z^2-\left(1-Y_z\right)^2+1\right]\Bigg\}\\
&=&-2\left(\cfrac{y_z}{2^{(2L-1)n-k}}-\cfrac{q_z+2^{(2L-1)n-k+1}\delta}{2^{(2L-1)n-k+1}}\right)^2+\cfrac{3}{2}+\cfrac{1}{2}\left(\cfrac{q_z+2^{(2L-1)n-k+1}\delta}{2^{(2L-1)n-k}}\right)^2.
\label{approx}
\end{eqnarray}
As a result, the expected value $\mathbb{E}_{b_z}[R(y_z,b_z)]$ of the reward is uniquely maximized when
\begin{eqnarray*}
y_z=\cfrac{q_z}{2}+2^{(2L-1)n-k}\delta\equiv y_{\rm max}.
\end{eqnarray*}
Since $|\delta|\le 2^{-f'(n)-(2L-1)n+k}$, $|y_{\rm max}-q_z/2|\le2^{-f'(n)}$.
This means that even in the approximation case, the rational server sends $\eta_z$ that is polynomially close to $y_{\rm max}$ and thus polynomially close to $q_z/2$.

Furthermore, from Eq.~(\ref{approx}), the expected value $\sum_{z\in\{0,1\}^k}\mathbb{E}_{b_z}[R(y_z,b_z)]$ of the total reward is lower-bounded by $3/2+O(1/2^{2(2L-1)n-k})$. Therefore, even in the approximation case, Protocol 1 satisfies the third condition in Sec.~\ref{I}.

\section{Appendix C: Estimation of $q_z$}
In this Appendix, we show that the quantum server can efficiently estimate $q_z$ with polynomial accuracy with high probability.
The server performs the following procedure:
\begin{enumerate}
\item Generate $U|0^n\rangle$, then measure the first $k$ qubits in the computational basis.
\item Output $X=1$ if the outcome in step 1 is $z$; otherwise, output $X=0$.
\item Repeat steps 1 and 2 $T$ times, then calculate $\eta_z=\sum_{i=1}^TX_i/T$.
\end{enumerate}
Using the Chernoff-Hoeffding bound~\cite{H63}, we immediately obtain
\begin{eqnarray*}
{\rm Pr}\left[|\eta_z-q_z|\ge\epsilon' \right]\le2e^{-2T{\epsilon'}^2}.
\end{eqnarray*}
Note that this procedure works for any $k$ (e.g., $k=O(\log{n})$ and $k=n$).

\section{Appendix D: The proof of Theorem~\ref{sparseapp}}
In this Appendix, we give the proof of Theorem~\ref{sparseapp}.\\
\noindent
{\it Proof.}
We use exactly the same argument as in Ref.~\cite{SN13}.
First, using the method in Appendix C with 
\begin{eqnarray*}
\epsilon'&\le&\cfrac{\epsilon}{l}\\
T&=&\cfrac{1}{2{\epsilon'}^2}\log{\cfrac{2(2t+\epsilon)}{\epsilon\delta}},
\end{eqnarray*}
it is possible to efficiently obtain $\eta_z$ for each $z\in \mathcal{L}$ such that $|\eta_z-q_z|\le\epsilon'$ with probability of at least $1-\epsilon\delta/(2t+\epsilon)$.
Therefore, we obtain a vector ${\bm\eta}_\mathcal{L}=(\eta_z : z\in \mathcal{L})$ such that $|\eta_z-q_z|\le\epsilon'$ for all $z\in \mathcal{L}$ and the list $\mathcal{L}$ includes all $z$ such that $q_z\ge\epsilon/t$, with probability of at least
\begin{eqnarray*}
\left[1-\cfrac{\epsilon\delta}{2t+\epsilon}\right]^{1+l}\ge 1-(1+l)\cfrac{\epsilon\delta}{2t+\epsilon}\ge 1-\delta.
\end{eqnarray*}
Setting $\eta_z$ to zero for all $z\notin \mathcal{L}$ and combining them with $\{\eta_z\}_{z\in \mathcal{L}}$, we define the $\lfloor 2t/\epsilon\rfloor$-sparse vector ${\bm\eta}\equiv(\eta_z : z\in\{0,1\}^n)$.

The task left is to show that $\sum_{z\in\{0,1\}^n}|\eta_z-q_z|\le 3\epsilon$ when $|\eta_z-q_z|\le\epsilon'$ for all $z\in \mathcal{L}$.
To show this, we define $A_t\subseteq\{0,1\}^n$ as a subset of $n$-bit strings such that $|A_t|=t$, and $q_z\ge q_y$ for any $z\in A_t$ and any $y\notin A_t$.
We also define $\tilde{\bm{Q}}\equiv(\tilde{q}_z : z\in\{0,1\}^n)$ as a vector where $\tilde{q}_z=q_z$ when $z\in A_t$; otherwise, $\tilde{q}_z=0$.
It is straightforward to show that for any $t$-sparse vector ${\bm v}=(v_z : z\in\{0,1\}^n)$, $\sum_{z\in\{0,1\}^n}|\tilde{q}_z-q_z|\le\sum_{z\in\{0,1\}^n}|v_z-q_z|$.
Therefore, from Definition~\ref{sparse},
\begin{eqnarray}
\label{hight}
\sum_{z\in\{0,1\}^n}|\tilde{q}_z-q_z|\le\epsilon.
\end{eqnarray}
From Eq.~(\ref{hight}) and $q_z\ge\epsilon/t\Rightarrow z\in \mathcal{L}$, which is the property of the list $\mathcal{L}$, when $|\eta_z-q_z|\le\epsilon'$ for all $z\in \mathcal{L}$,
\begin{eqnarray*}
\sum_{z\in\{0,1\}^n}|\eta_z-q_z|&=&\sum_{z\in \mathcal{L}}|\eta_z-q_z|+\sum_{z\notin \mathcal{L}}q_z\\
&\le&l\epsilon'+\sum_{z\notin \mathcal{L}}|q_z-\tilde{q}_z|+\sum_{z\notin \mathcal{L}}\tilde{q}_z\\
&\le&\epsilon+\epsilon+|A_t|\cfrac{\epsilon}{t}\\
&=&3\epsilon.
\end{eqnarray*}
\hspace{\fill}$\blacksquare$

\section{Appendix E: The proof of Theorem~\ref{sparsesamp}}
In this Appendix, we give the proof of Theorem~\ref{sparsesamp}.\\
\noindent
{\it Proof.} The argument is the same as in Ref.~\cite{SN13}. In this proof, we assume that $|\eta_z-q_z|\le\epsilon'$ for all $z\in \mathcal{L}$.
From the proof of Theorem~\ref{sparseapp}, this assumption is true with probability of at least $1-\delta$.
We define $p_z$ as $\eta_z/(\sum_{z\in\{0,1\}^n}\eta_z)$.
Since $\{\eta_z\}_{z\in\{0,1\}^n}$ is $\lfloor 2t/\epsilon\rfloor$-sparse, and $t$ and $1/\epsilon$ are polynomials in $n$, the calculation of the denominator can be performed in classical polynomial time.
Since $\sum_{z\in\{0,1\}^n}|\eta_z-q_z|\le 3\epsilon$, we obtain $1-3\epsilon\le\sum_{z\in\{0,1\}^n}\eta_z\le 1+3\epsilon$.
Therefore,
\begin{eqnarray*}
\sum_{z\in\{0,1\}^n}|p_z-q_z|&=&\sum_{z\in\{0,1\}^n}\cfrac{\left|\eta_z-\left(\sum_{z'\in\{0,1\}^n}\eta_{z'}\right)q_z\right|}{\sum_{z'\in\{0,1\}^n}\eta_{z'}}\\
&\le&\sum_{z\in\{0,1\}^n}\cfrac{\left|\eta_z-\left(\sum_{z'\in\{0,1\}^n}\eta_{z'}\right)q_z\right|}{1-3\epsilon}\\
&\le&\sum_{z\in\{0,1\}^n}\cfrac{\left|\eta_z-q_z\right|}{1-3\epsilon}+\sum_{z\in\{0,1\}^n}\cfrac{\left|1-\sum_{z'\in\{0,1\}^n}\eta_{z'}\right|q_z}{1-3\epsilon}\\
&\le&\cfrac{3\epsilon+3\epsilon}{1-3\epsilon}\\
&\le&12\epsilon,
\end{eqnarray*}
where we have used $\epsilon\le 1/6$ to derive the last inequality.
\hspace{\fill}$\blacksquare$

\section{Appendix F: The proof of Theorem~\ref{nogo}}
In this Appendix, we give the proof of Theorem~\ref{nogo}\\
\noindent
{\it Proof.} We use the same argument used in Ref.~\cite{CMS18}.
The prover and the verifier perform the following procedure:
\begin{enumerate}
\item The prover and the verifier simulate the constrained RDQC protocol except for paying the reward.
\item The verifier calculates the value of the reward $R$.
\item If $o=1$, the verifier accepts with probability $R/c$. Otherwise, the verifier rejects.
\end{enumerate}
First, assume that $x\in L$. From Definition~\ref{RDQCRDQC}, if the prover decides a strategy $s$ following the distribution $\mathcal{D}'_{\rm YES}$, the verifier's acceptance probability $p_{\rm acc}$ is
\begin{eqnarray}
\label{completeness}
p_{\rm acc}=\cfrac{\mathbb{E}_{s\sim\mathcal{D}'_{\rm YES}}[R(s,x)]}{c}\ge\cfrac{c_{\rm YES}}{c}.
\end{eqnarray}
Then assume that $x\notin L$. If the prover takes a strategy $s_0$ that makes the verifier set $o=0$, the acceptance probability is $p_{\rm acc}=0$. On the other hand, if the prover takes a strategy $s_1$ that makes the verifier set $o=1$,
\begin{eqnarray}
\label{soundness}
p_{\rm acc}\le\cfrac{{\rm max}_{s_1\in S_{\rm incorrect},x\notin L}\mathbb{E}[R(s_1, x)]}{c}.
\end{eqnarray}
From Eqs.~(\ref{add}), (\ref{completeness}), and (\ref{soundness}), the completeness-soundness gap $(c_{\rm YES}-{\rm max}_{s_1\in S_{\rm incorrect},x\notin L}\mathbb{E}[R(s_1, x)])/c$ is lower bounded by $1/(cf(|x|))>1/{\rm poly}(|x|)$.
\hspace{\fill}$\blacksquare$

\end{document}